\documentclass{article}
\usepackage[T1]{fontenc}
\usepackage[latin9]{inputenc}
\usepackage{geometry}
\usepackage{verbatim}
\usepackage{amsmath}
\usepackage{amssymb}
\usepackage{graphicx}
\pdfoutput=1
\makeatletter

\providecommand{\tabularnewline}{\\}

\let\jnl@style=\rm
\def\ref@jnl#1{{\jnl@style#1}}

\def\aj{\ref@jnl{AJ}}                   
\def\actaa{\ref@jnl{Acta Astron.}}      
\def\araa{\ref@jnl{ARA\&A}}             
\def\apj{\ref@jnl{ApJ}}                 
\def\apjl{\ref@jnl{ApJ}}                
\def\apjs{\ref@jnl{ApJS}}               
\def\ao{\ref@jnl{Appl.~Opt.}}           
\def\apss{\ref@jnl{Ap\&SS}}             
\def\aap{\ref@jnl{A\&A}}                
\def\aapr{\ref@jnl{A\&A~Rev.}}          
\def\aaps{\ref@jnl{A\&AS}}              
\def\azh{\ref@jnl{AZh}}                 
\def\baas{\ref@jnl{BAAS}}               
\def\bac{\ref@jnl{Bull. astr. Inst. Czechosl.}}
\def\caa{\ref@jnl{Chinese Astron. Astrophys.}}
\def\cjaa{\ref@jnl{Chinese J. Astron. Astrophys.}}
\def\icarus{\ref@jnl{Icarus}}           
\def\jcap{\ref@jnl{J. Cosmology Astropart. Phys.}}
\def\jrasc{\ref@jnl{JRASC}}             
\def\memras{\ref@jnl{MmRAS}}            
\def\mnras{\ref@jnl{MNRAS}}             
\def\na{\ref@jnl{New A}}                
\def\nar{\ref@jnl{New A Rev.}}          
\def\pra{\ref@jnl{Phys.~Rev.~A}}        
\def\prb{\ref@jnl{Phys.~Rev.~B}}        
\def\prc{\ref@jnl{Phys.~Rev.~C}}        
\def\prd{\ref@jnl{Phys.~Rev.~D}}        
\def\pre{\ref@jnl{Phys.~Rev.~E}}        
\def\prl{\ref@jnl{Phys.~Rev.~Lett.}}    
\def\pasa{\ref@jnl{PASA}}               
\def\pasp{\ref@jnl{PASP}}               
\def\pasj{\ref@jnl{PASJ}}               
\def\rmxaa{\ref@jnl{Rev. Mexicana Astron. Astrofis.}}%
\def\qjras{\ref@jnl{QJRAS}}             
\def\skytel{\ref@jnl{S\&T}}             
\def\solphys{\ref@jnl{Sol.~Phys.}}      
\def\sovast{\ref@jnl{Soviet~Ast.}}      
\def\ssr{\ref@jnl{Space~Sci.~Rev.}}     
\def\zap{\ref@jnl{ZAp}}                 
\def\nat{\ref@jnl{Nature}}              
\def\iaucirc{\ref@jnl{IAU~Circ.}}       
\def\aplett{\ref@jnl{Astrophys.~Lett.}} 
\def\apspr{\ref@jnl{Astrophys.~Space~Phys.~Res.}}
\def\bain{\ref@jnl{Bull.~Astron.~Inst.~Netherlands}} 
\def\fcp{\ref@jnl{Fund.~Cosmic~Phys.}}  
\def\gca{\ref@jnl{Geochim.~Cosmochim.~Acta}}   
\def\grl{\ref@jnl{Geophys.~Res.~Lett.}} 
\def\jcp{\ref@jnl{J.~Chem.~Phys.}}      
\def\jgr{\ref@jnl{J.~Geophys.~Res.}}    
\def\jqsrt{\ref@jnl{J.~Quant.~Spec.~Radiat.~Transf.}}
\def\memsai{\ref@jnl{Mem.~Soc.~Astron.~Italiana}}
\def\nphysa{\ref@jnl{Nucl.~Phys.~A}}   
\def\physrep{\ref@jnl{Phys.~Rep.}}   
\def\physscr{\ref@jnl{Phys.~Scr}}   
\def\planss{\ref@jnl{Planet.~Space~Sci.}}   
\def\procspie{\ref@jnl{Proc.~SPIE}}   

\makeatother

\begin{document}

\title{Normal type Ia supernovae from disruptions \\
of hybrid He-CO white-dwarfs by CO white-dwarfs}

\author{Hagai B. Perets$^{1}$\thanks{Contributed equally}, Yossef Zenati$^{*}$$^{1}$,
Silvia Toonen$^{2,1}$ \& Alexey Bobrick$^{3}$}
\maketitle
\begin{center}
{\small{}$^{1}$Technion - Israel Institute of Technology, Physics
department, Haifa Israel 3200002}{\small\par}
\par\end{center}

\begin{center}
{\small{}$^{2}$Institute of Gravitational Wave Astronomy, School
of Physics and Astronomy, University of Birmingham, Birmingham B15
2TT, UK}{\small\par}
\par\end{center}

\begin{center}
{\small{}$^{3}$Lund University, Department of Astronomy and Theoretical
physics, Box 43, SE 221-00 Lund, Sweden}{\small\par}
\par\end{center}
\begin{abstract}
Type Ia supernovae (SNe) are thought to originate from the thermonuclear
explosions of carbon-oxygen (CO) white dwarf (WD) stars\cite{2014ARA&A..52..107M,Liv+18,Wan+18}.
They produce most of the iron-peak elements in the universe, and bright
type Ia SNe serve as important ``standard candle'' cosmological
distance indicators. The proposed progenitors of standard type Ia
SNe have been studied for decades, and can be, generally, divided
into explosions of CO WDs accreting material from stellar non-degenerate
companions (single-degenerate; SD models), and those arising from
the explosive interaction of two CO WDs (double-degenerate; DD models).
However, current models for the progenitors of such SNe fail to reproduce
the diverse properties of the observed explosions, nor do they explain
the inferred rates and the characteristics of the observed populations
of type Ia SNe and their expected progenitors\cite{2014ARA&A..52..107M,Liv+18,Wan+18}.
Here we show that the little studied mergers of CO-WDs with hybrid
Helium-CO (He-CO) WDs can provide for a significant population of
the normal type Ia SNe. Population synthesis studies showed that a
large fraction (15-30$\%$) of all WD-WD mergers could involve a unique
type of \emph{Hybrid He-CO WD}s\cite{Liu+17,Yun+17,Zen+18a}. Although
it was suggested that they may play a role as DD progenitors of normal
or peculiar type Ia SNe\cite{Dan+14,Dan+15,Liu+17,Yun+17,Zen+18a},
no detailed explosion models with observable light-curve and spectra
predictions had ever tested this possibility. Here we use detailed
thermonuclear-hydrodynamical and radiative-transfer models to show
that a wide range of mergers of CO WDs with hybrid He-CO WDs (see
\cite{Ibe+85,Zen+18a} and references therein) can give rise to normal
type Ia SNe. We find that such He-enriched mergers give rise to explosions
for which the synthetic light-curves and spectra resemble those of
observed type Ia SNe, and in particular they can produce a wide range
of peak-luminosities, ${\rm M_{B}(M_{R})\sim}$$-18.4$ to $-19.2$
($\sim-18.5$ to $-19.45$), consistent with those observed for normal
type Ia SNe. Moreover, our population synthesis models show that,
together with the contribution from mergers of massive double CO-WDs
(producing the more luminous SNe), they can potentially reproduce
the full range of type Ia SNe, their rate and delay-time distribution.
Mergers of hybrid He-CO WDs can therefore play a key role in explaining
the origin of type Ia SNe, serve to study their detailed composition
yields, and potentially probe the systematics involved in type Ia
SNe measurements of the cosmological parameters of the universe. 
\end{abstract}
Both the SD and DD progenitor models proposed for type Ia SNe encounter
major challenges. Most of the population synthesis studies suggest
that the estimated rates of type Ia SNe from the SD channel are too
low in comparison with the rates inferred from observations, and the
explosions occur too early as to explain the existence SNe in old-environments\cite{2014ARA&A..52..107M}.
Moreover, surveys searching for accreting WD progenitors failed to
detect sufficient numbers of such progenitors\cite{Dis+10,Joh+14,2014ARA&A..52..107M}.
Furthermore, in several cases stringent limits were derived on the
possible existence of a stellar mass-donor companion expected in this
scenario\cite{Gra+15}. The estimated rates from the DD scenario can
be consistent with observations only if at least $\sim14\%$ of all
double-WD (DWD) mergers produce type Ia SNe\cite{Too+12,Mao+h17}.
In particular, these rate constraints require the more frequent binary-mergers,
involving low-mass ($<\sim0.85$ M$_{\odot}$) WDs to give rise to
such SNe. However, all merger-models (as well as dynamical detonation
of low-mass He-shell models, e.g. \cite{She+18}) of such low-mass
CO and/or He WDs fail to explode or produce very faint SNe in models
suggested to-date. Moreover, the few successful explosion models of
higher mass WDs (CO-CO WD mergers with total combined mass $\gtrsim1.9\,{\rm M_{\odot})}$
which light-curve and spectra have been modeled in detail either give
rise to peculiar SNe and/or produce only high luminosity and slow
evolving SNe, while failing to reproduce the diverse characteristics
of lower luminosity and/or slower-evolving population of normal type
Ia SNe (see \cite{Pak17} for an overview). Other models such as the
core-degenrate models and collisions in triple systems have not yet
modelled in depth, and/or their rates and delay time distribution
are inconsistent with those of normal Ia SNe \cite{Ilk+12,Zho+15,Kus+13,Too+18b}.
Here we show that DWD-mergers involving a \emph{different} type of
WDs, namely \emph{hybrid He-CO WDs}, that were little explored before,
can reproduce the detailed properties of a wide range of normal type
Ia SNe, their characteristic rates and their delay-time distributions.

WDs formed through stellar evolution of a single star at the current
age of the universe are CO WDs in the mass range ${\rm \sim0.50-1.05M}$
and O-Ne WDs in the range $\sim1.05-1.38).$ However, binary evolution
can give rise to WDs with very different properties. In interacting
binaries each of the stellar components may fill its Roche lobe, and
may be stripped of part of its hydrogen and/or Helium-rich envelope
during its evolution on the red giant branch or the asymptotic giant
branch stage. Such altered evolution can give rise to qualitatively
different evolution and the formation of present day WDs with a significant
fraction of He mass. The evolution and final outcomes of the binary
evolution strongly depend on the initial conditions: the mass of the
stellar components and their initial separation. In particular, WDs
of masses lower than 0.45 ${\rm M_{\odot}}$ are typically thought
to be Helium (He)-WDs formed through this channel \cite{Ibe+85,Han+02,Nel+01,Ist+16,Zha+18}.
However, the complex binary evolution channel can give rise to \emph{hybrid-WDs},
composed of significant fractions of both CO and He. Such white dwarfs
descend from stars which fill their Roche lobes in the stage of hydrogen
burning in a shell, become hot sub-dwarfs in the He-burning stage,
but do not experience envelope expansion after the formation of a
degenerate carbon-oxygen core \cite{Ibe+85,Nel+00,Zen+18a}. Such
hybrid WDs reside in the mass range of $0.4-0.72$ ${\rm M_{\odot}}$
and contain a He-envelope containing $\sim2-20\%$ of the WD-mass
(with the rest composed of a CO core)\cite{Zen+18a}. Though the mergers
of CO, He, and O-Ne WDs (and their various combinations) have been
explored in detail, hybrid He-CO WDs have been little explored, and
their light-curve and spectra have never been modeled nor directly
compared with observations. This missing piece in our understanding
of double-WD mergers and their outcomes is of particular interest
given that mergers involving such hybrid-WDs are expected to comprise
a significant fraction of $\sim15-30\%$ of all DWD-mergers as we
discuss below (and consistent with previous studies\cite{Liu+17,Yun+17}).
As we show below these can produce successful type Ia SNe, and the
phase-space of DWD-merger combinations involving hybrid He-CO WDs
can reproduce the detailed characteristics and demographics of most
of the observed type Ia SNe. 

When the densities of two merging WDs sufficiently differ, the less
dense and less massive WD is can begin transfer mass to its companion,
and later be tidally disrupted by the more massive and compact WD
before the WDs attain direct physical contact. The debris of the disrupted
WD might then form an accretion disk around the more massive WD. Naturally,
in order to fully model this complex evolution one requires a 3D simulation.
However, 3D simulations are numerically highly prohibiting, and provide
poor resolution, not sufficient to correctly resolve the relevant
underlying physical processes involved. Therefore, in order to enable
efficient modeling of a wide phase-space of hybrid He-CO WD mergers
with CO WDs we use an alternative route, making use of 2D models,
but still capturing the important 3D aspects of the merger, as we
describe in the following. Naturally, this approach can not capture
potential nuclear burning that occurs during the early phases of the
mergers and/or during the disruption of the WD. Such processes which
might affect the evolution (e.g. in the suggested models of the detonation
of very thin He layer possibly dynamically accreted and/or detonated
in these early stages\cite{She+18}); nor can it consistently explore
the (small fraction of) mergers of comparable-mass WDs in which cases
a violent merger is expected to occur\cite{Pak17}, not leading to
the formation of an accretion disk, as assumed in the 2D models. These
aspects will be explored in the future via full 3D simulations. 

Although the debris disk from the disrupted WD can be initially clumpy
and/or asymmetric, it can rapidly evolve into a relatively symmetric
accretion disk around the more massive WD before any significant nuclear
burning occurs\cite{Dan+14,VanrossumWollaeger16}, though spiral mode
instabilities may occur in mergers of comparable mass massive WDs\cite{kas+17}.
We therefore model the mergers starting only following the formation
of a symmetric debris disk around the more massive WD in our simulations,
similar to the approach used by us and others \cite{Fer+13,Dan+15,Zen+18b}
to model mergers of WDs with WDs/neutron-stars (NSs). Such cylindrical
symmetry of the disk and the central accreting WD allows us to model
the merger in 2D. Such models are highly advantageous as they allow
for high resolution simulations, with relatively little numerical
expense in comparison with 3D simulation, but they still capture most
of the important multi-dimensional aspects of the merger.

We explored a wide range of combinations of WD mergers involving a
hybrid He-CO-{}-WD merging with a more massive CO WD (see Table \ref{tab:initial-final}).
We considered mergers with total combined masses ranging between $\sim1.2$
${\rm M_{\odot}}$and $1.75$ ${\rm {\rm M_{\odot}}}$. In all cases
the lower-mass hybrids can be assumed to be fully disrupted by the
CO WDs. All of these models gave rise to a full detonation of the
CO WD and the nuclear burning of the debris disk (as we discuss below).
Models with total mass above $\sim1.15$ ${\rm M_{\odot}}$ (all the
models in Table \ref{tab:initial-final}) produced SNe resembling
normal type Ia SNe. We also modeled inverse cases where the hybrids
disrupted the CO WDs, mergers of double hybrid WDs and cases with
lower total masses. We find that these latter models do not produce
normal type Ia SNe, but can explain several of the relatively frequent
peculiar types of thermonuclear SNe (to be discussed elsewhere). 

All of the WD profiles used in our models were produced using the
stellar evolution code MESA\cite{Pax+15}; using the existing inlists
in the MESA website for the models of CO WDs. The hybrid WDs were
taken from our detailed MESA models for the formation of hybrid He-CO
WDs\cite{Zen+18a}. The 2D hydrodynamical-thermonuclear simulations
of the mergers were made using the publicly available FLASH v4.5 code
\cite{2000ApJS..131..273F}, and following the same methods we previously
employed\cite{Zen+18b}. We applied a burning limiter approach following
\cite{2013ApJ...778L..37K}, as to quench artificial burning (see
Methods \ref{subsec:FLASH+supernu} for details). Each of our simulations
included a large number (5k-10k) of tracer test-particles, used for
post-processing analysis with a large nuclear network (125 isotopes),
applying the MESA-TORCH module, and followed by radiative transfer
modeling using the openly available SuperNu code\cite{Wol+13,wollaegervanrossum14},
as to produce the multi-band light-curves and detailed spectra of
the simulated SNe. The density, composition and temperature profiles
of the central CO WDs were mapped from the 1D MESA models into the
2D FLASH simulations. The hybrid He-CO WDs were modeled as self-consistent
accretion disks following the same procedure used by us earlier\cite{Zen+18b},
where we assume the disk composition is fully mixed (consistent with
SPH results), and follows the composition of the He-CO WDs as found
in our stellar-evolution He-CO-WD models\cite{Zen+18a}. These conditions
are generally consistent with the results of 3D SPH simulations of
such mergers\cite{Dan+14}. The position of the inner radius of the
disk at the beginning of the simulation is expected to be of the order
of the tidal disruption radius, but the exact position is not known
a priori, and we therefore explored a variety of models, including
cases with inner disk radii of $r_{in}=R_{t}$ and a few cases where
we considered $r_{in}=0.8R_{t}$, where $R_{t}$ is the tidal radius
$\sim{\rm (M_{CO}^{WD}/M_{CO-He}^{WD}})^{1/3}{\rm R_{CO-He}^{WD}}$
and ${\rm M^{WD}}$and ${\rm {\rm R^{WD}}}$ are the masses and radii
of the WDs. The modeling and evolution of the debris disk, including
nuclear burning and viscous evolution follows the same procedures
applied by us previously\cite{Zen+18b}. More detailed discussion
of these and other numerical aspects can be found in the Methods section.

\begin{figure*}
\includegraphics[scale=0.22]{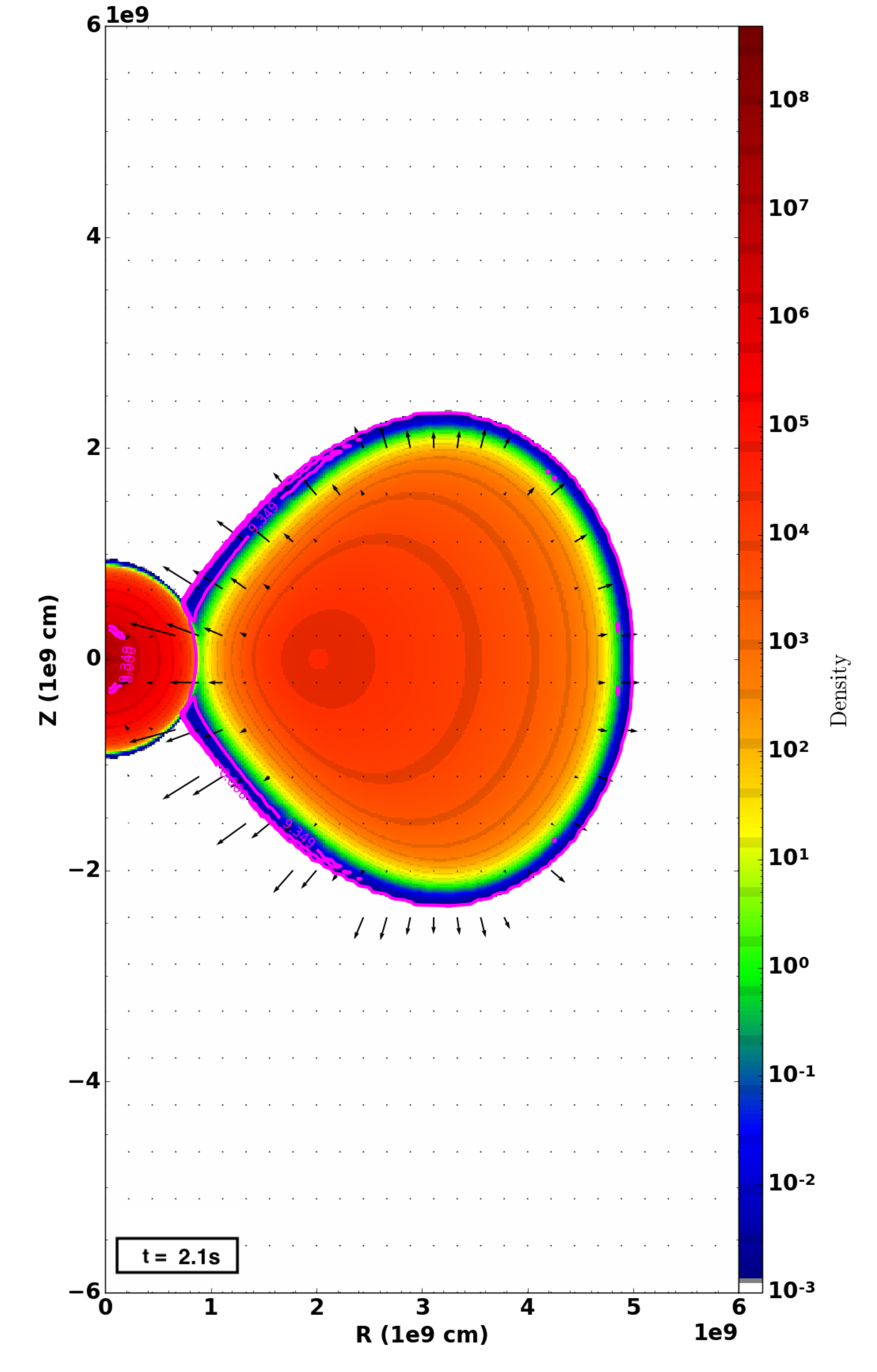}\includegraphics[scale=0.22]{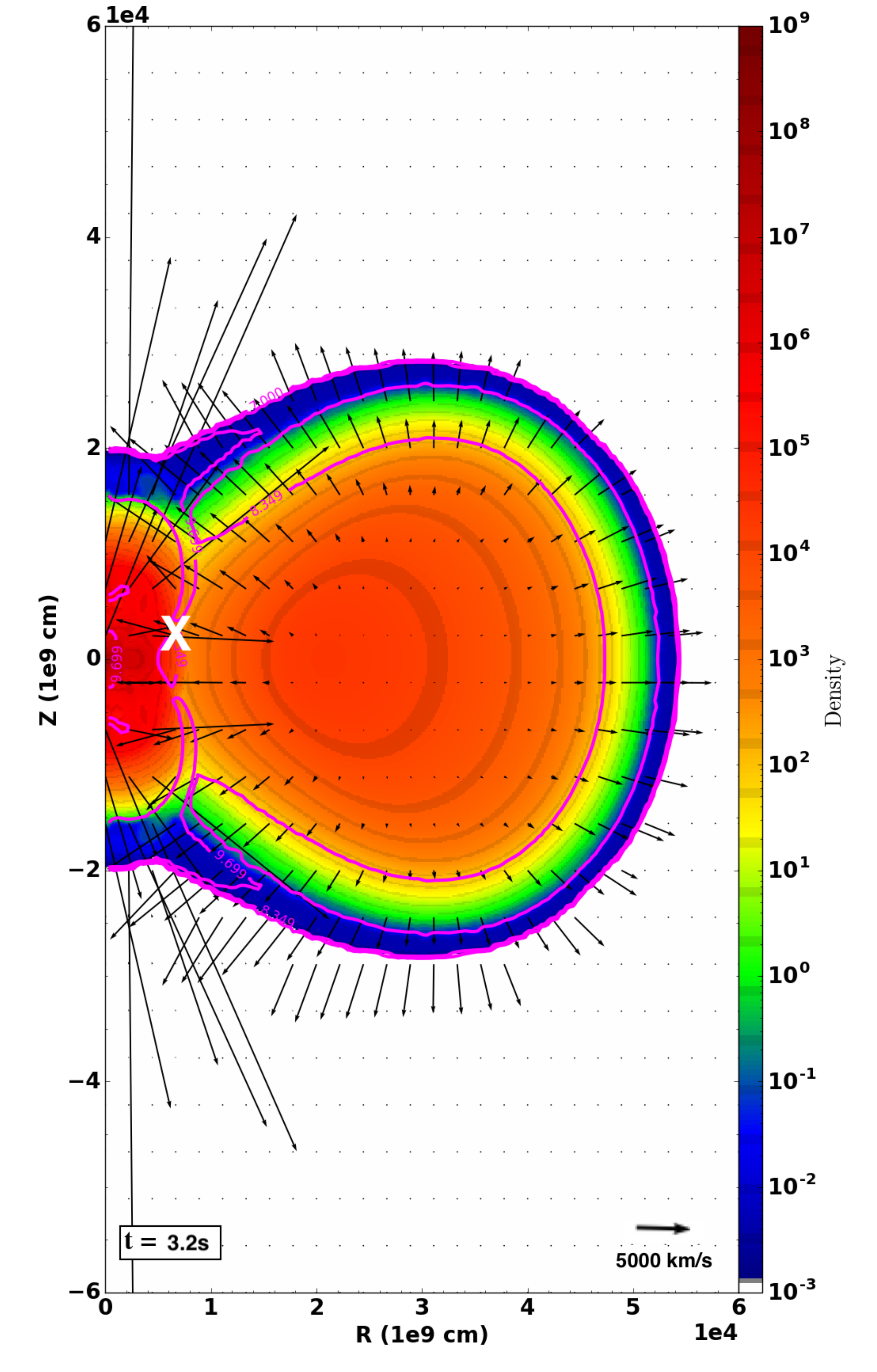}\includegraphics[scale=0.22]{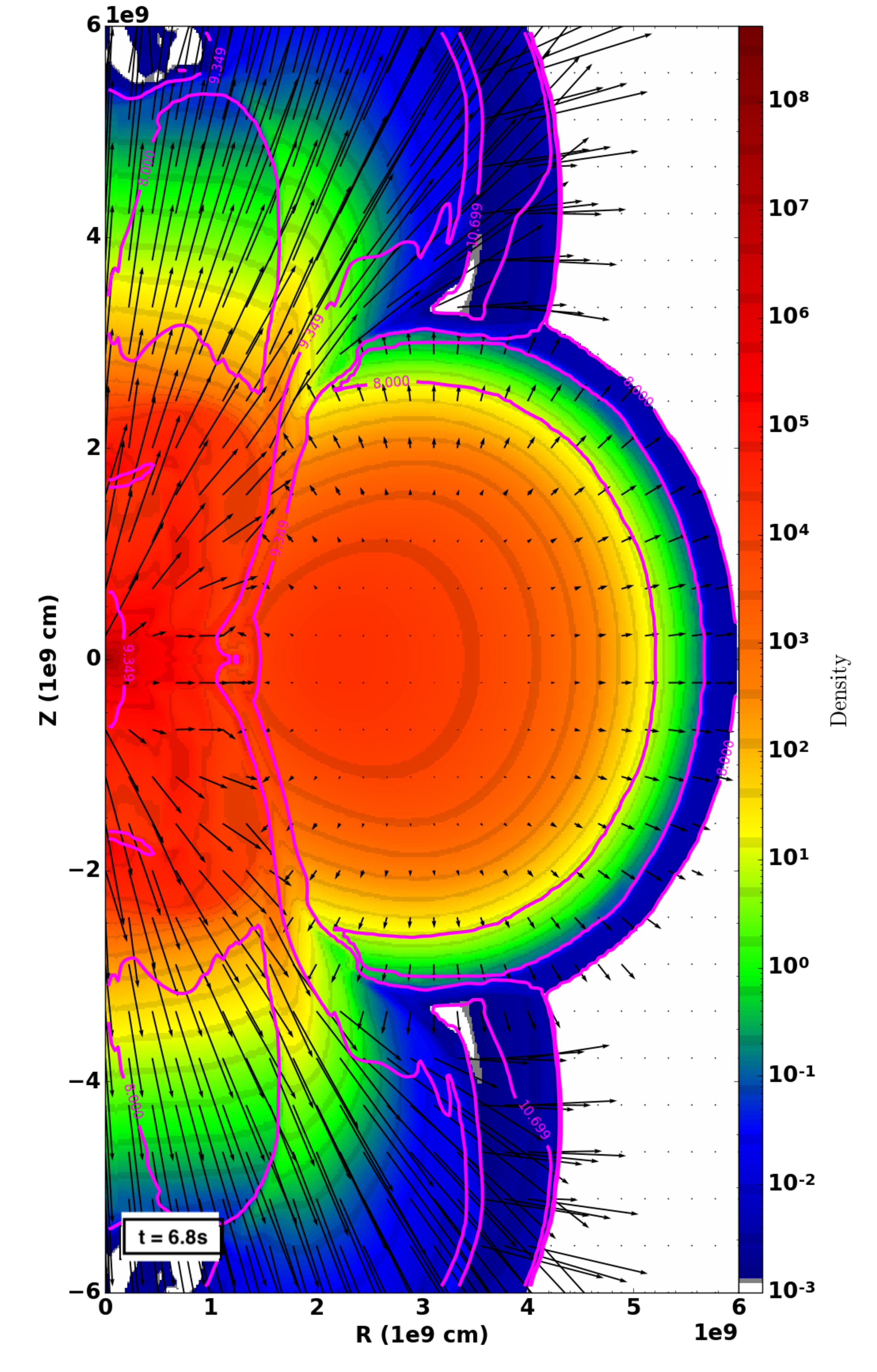}

\includegraphics[scale=0.22]{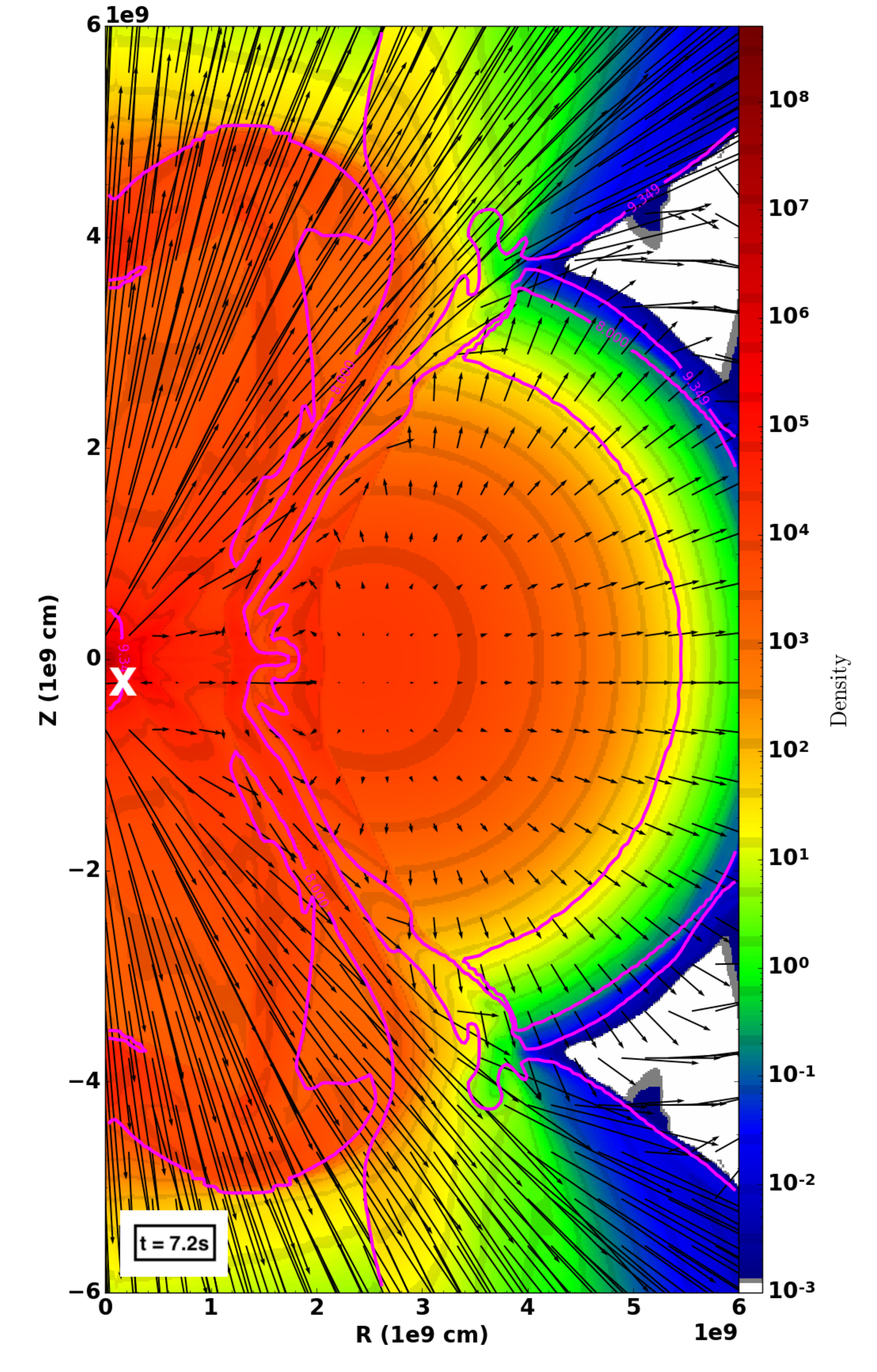}\includegraphics[scale=0.22]{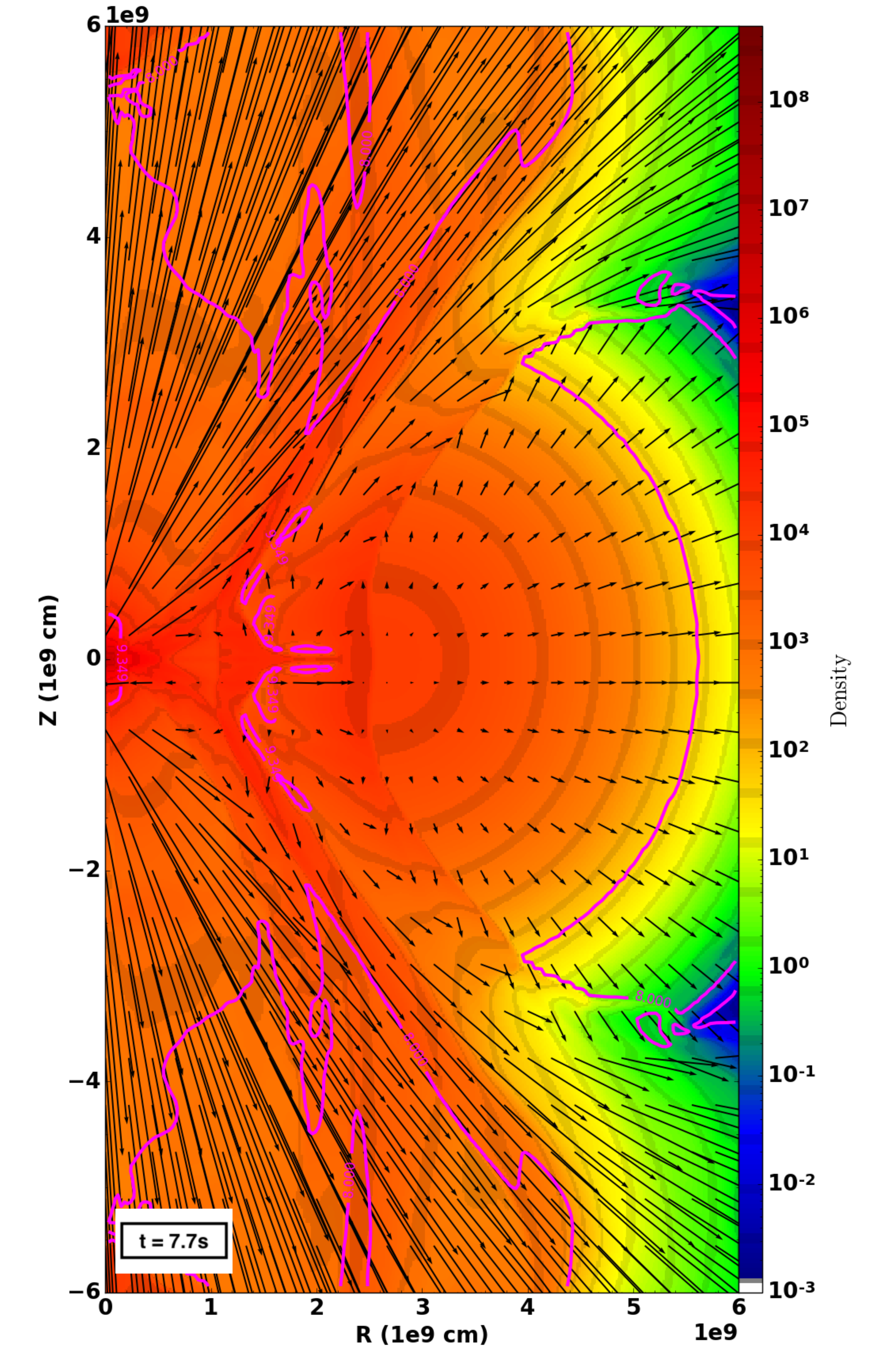}\includegraphics[scale=0.22]{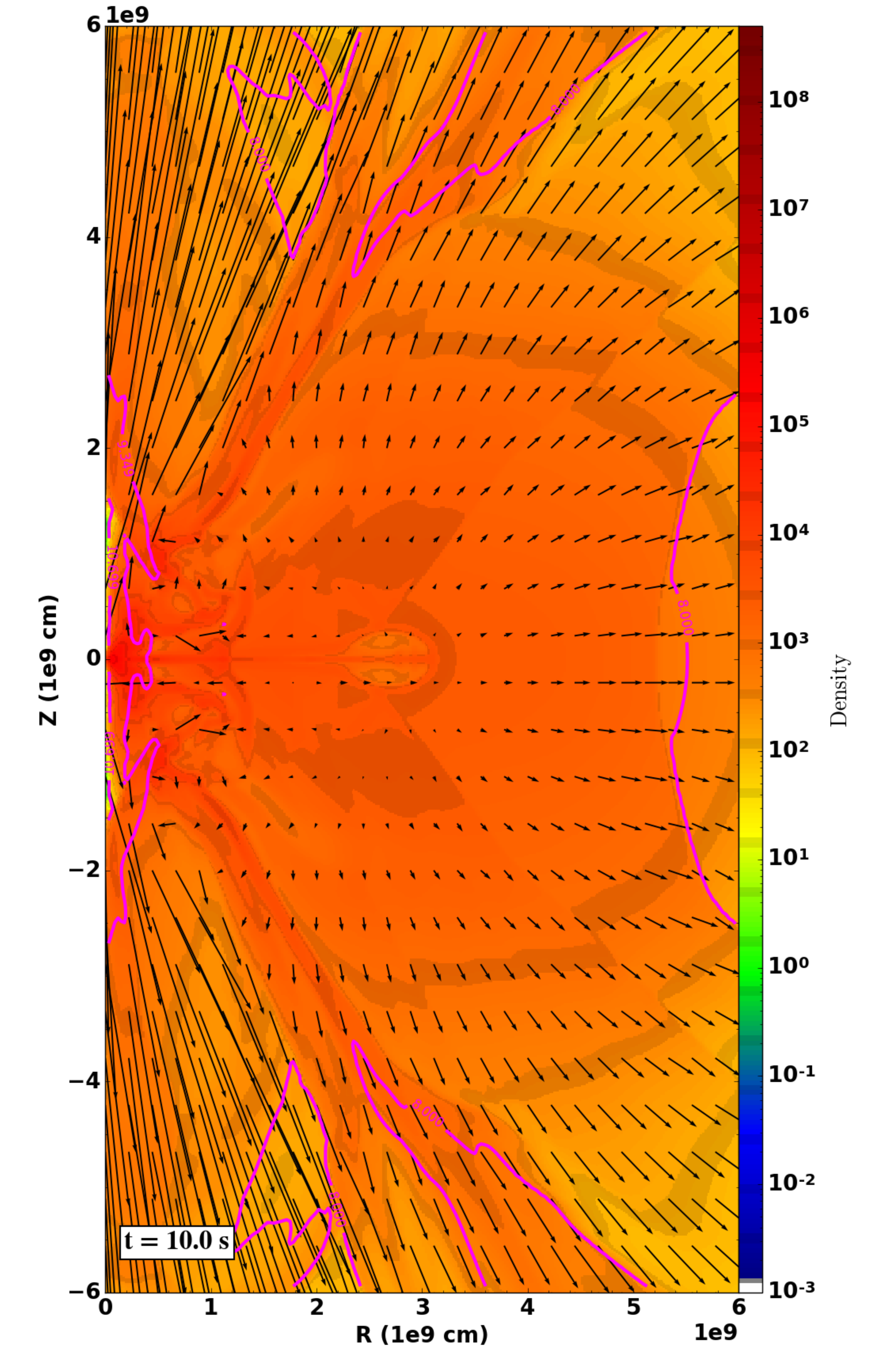}

\caption{\label{fig:disk-evolution-wd}{\footnotesize{}The evolution of the
debris disk from a 0.74 ${\rm M_{\odot}}$HeCO WD accreting on a 0.9
${\rm M_{\odot}}$ CO WD, following the disruption of the HeCO WD.
Each panel shows the (color coded) density distribution and velocity
vectors (black arrows; see top right panel showing 5000 km s$^{-1}$
arrow for calibration) at different times. Magenta contours correspond
to the temperature. As can be seen the debris disk evolves viscously
both inwards and outwards. Following the accretion on the surface
of CO WD, nuclear burning ensues on at the contact regions. At $t=3.2$
s a He-detonation occurs in He-mixed debris accreting on the WD (the
initial detonation point is marked with a white $\times$ (at $r=8.3\times10^{8},z=1.2\times10^{8}\,{\rm cm}$).
The burning front propagates into the interior of the CO WD interior,
compresses and heats it, and eventually at $t=7.2$ s a CO-detonation
is initiated close to the central part, at $r=4.7\times10^{8},z=-0.94\times10^{8}\,{\rm cm}$
from the center of the CO WD (the initial detonation point is marked
with a white $\times$). At the time of the CO detonation the densities
in the CO WD core are already significantly elevated due to the earlier
shock-compression, allowing for an efficient burning of the CO core
upto $^{56}{\rm Co}$, explaining the $^{56}{\rm Ni}$ mass produced
in these mergers, even for relatively low-mass CO WDs, which densities
were otherwise too-low as to allow for significant $^{56}{\rm Ni}$
in the absence of the earlier compression due to the He-detonation. }}
\end{figure*}

\begin{figure}
\includegraphics[scale=0.3]{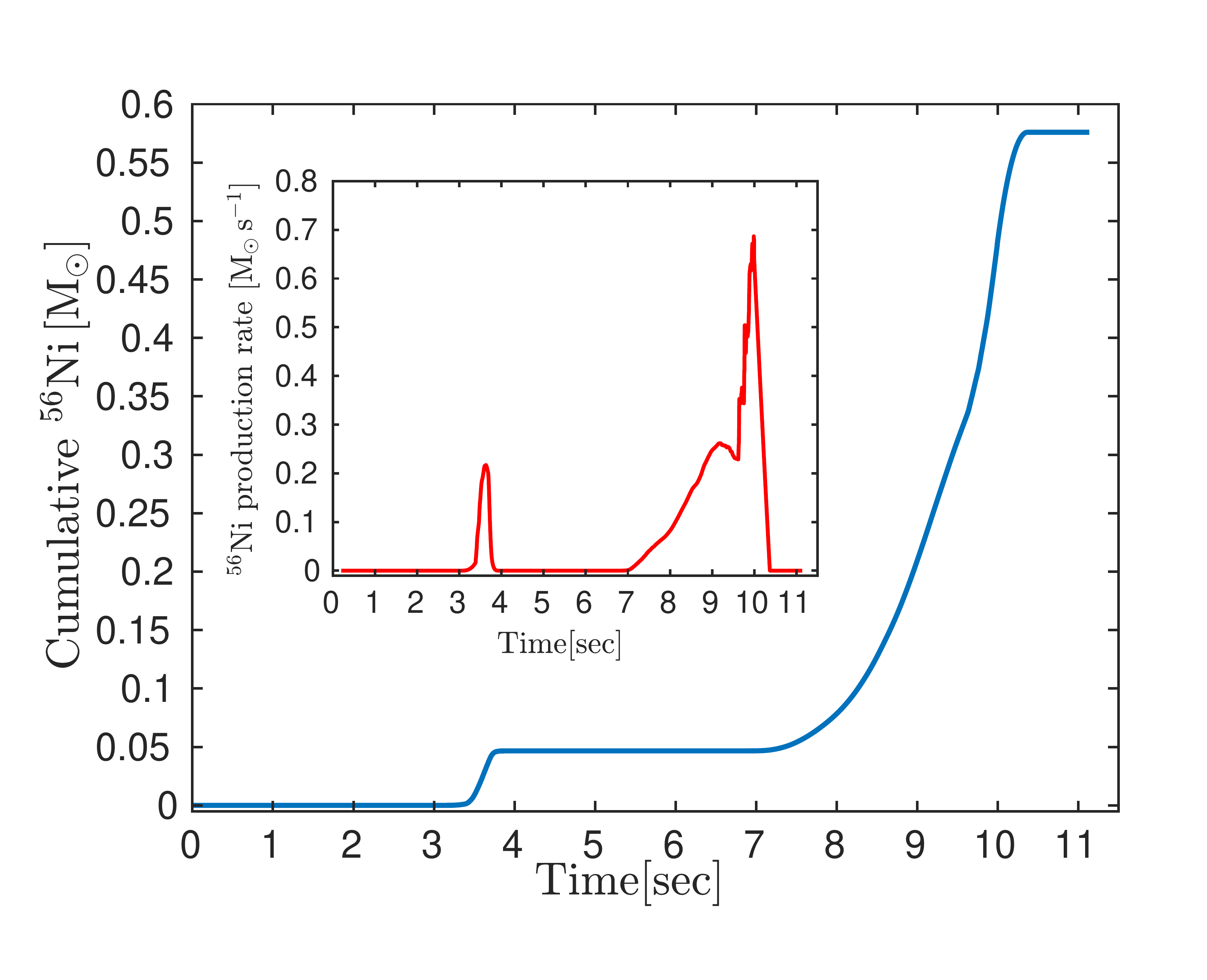}\caption{{\small{}\label{fig:56Ni-production}}{\footnotesize{}The production
rate of $^{56}{\rm Co}$ (to later decay to $^{56}$Ni) as a function
of time (for the same model shown in Fig. \ref{fig:disk-evolution-wd}).
The increase in production following the early He-detonation at 3
s and later the CO-detonation at 7 s can be clearly seen.}}
\end{figure}

Although the evolution of each of the CO - He-CO WD-WD merger models
shown here and its final outcomes depend on the specific initial conditions,
the overall evolution of the mergers follow a very similar behavior,
and we therefore focus on one example, shown in Fig. \ref{fig:disk-evolution-wd}.
Initially the disk evolves radially through viscous evolution; and
angular momentum exchange lead the inner regions of the disk to spread
inwards and heat up, while the outer regions expand outwards. Material
then accretes inwards through the (horizontal) central parts of the
disk and outflows of material ensues after a few seconds. The outflows
are ejected at a wide angle from the innermost central parts, as gravitational
accretion energy is converted into heat and kinetic energy fueling
the outflows. Once the disk material comes in contact and accretes
onto the central WD (at $\sim3$ s), nuclear burning ensues on the
surface of the CO WD, which further heats up the inner disk and the
outer layers of the central CO WD. At that point, at time $t=3.2$
a He-detonation ensues (see Fig. \ref{fig:disk-evolution-wd}) in
the He-mixed debris accreting on the CO WD, producing $\sim0.05$
${\rm M_{\odot}}$of $^{56}{\rm Ni}$ (see Fig. \ref{fig:56Ni-production}).
The burning-front then propagates inwards inside the WD (and outwards
in the debris disk) compresses the CO WD (see Methods \ref{subsec:shock}),
which then catalyzes a second, CO-detonation in the dense inner part
of the CO WD, leading to the full explosion of the system (we find
the CO-detonation condition is first triggered at $t\sim7.2$ s at
$r=4.7\times10^{8},z=-0.94\times10^{8}\,{\rm cm}$ ; see detonation
point marked in the figure). The detonation shock-wave propagates
and incinerates the CO WD and then burns through the now already more
spherically-symmetric debris-envelope composed of the He-CO WD debris.
As the burning front propagates into the outer, colder and less dense
regions, the burning efficiency rapidly drops, but only after already
burning nearly all of the He-mass in the debris-envelope (with only
$\sim10^{-5}-10^{-3}$${\rm M_{\odot}}$ of unburned He remaining),
and the final production of $\sim0.56$ ${\rm M_{\odot}}$ of $^{56}{\rm Ni}$
(see fig. \ref{fig:56Ni-production}) in the models shown (and a wider
range between $\sim0.5-0.6$ M$_{\odot}$ for all of the models explored;
see Table \ref{tab:initial-final}) and significant yields of iron-peak
and intermediate elements. 

\begin{figure}
\includegraphics[scale=0.3]{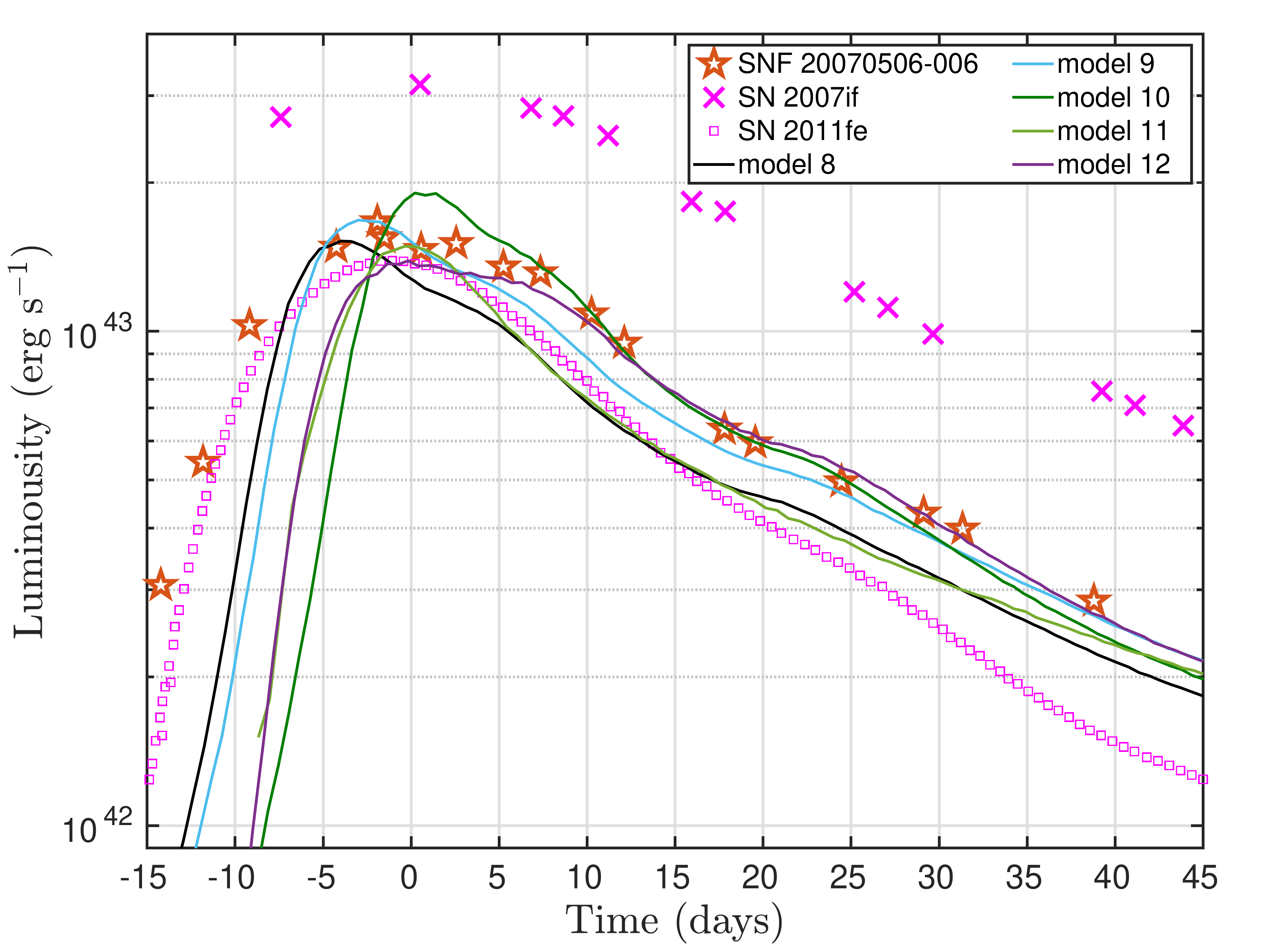}\includegraphics[scale=0.3]{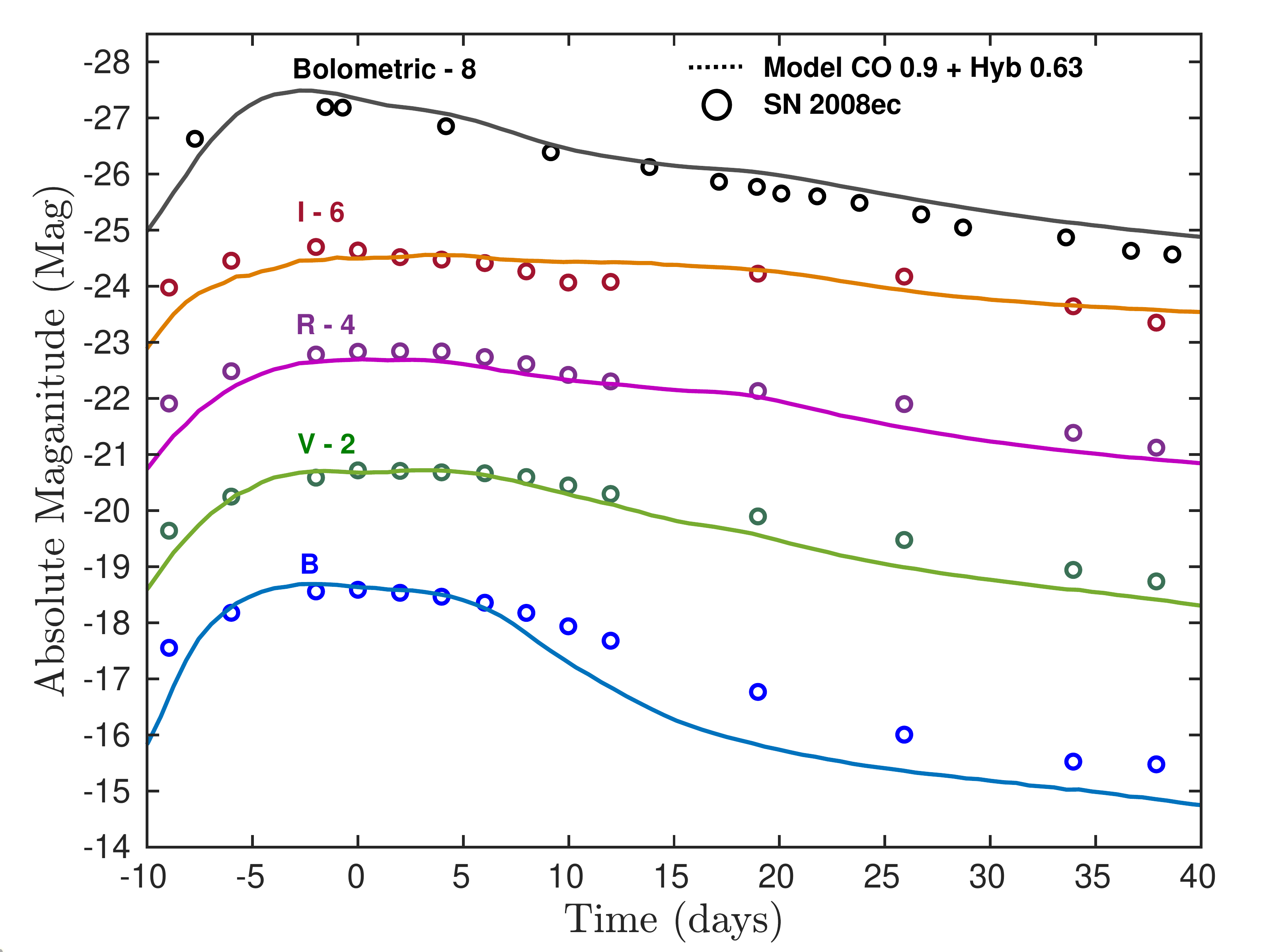}

\includegraphics[scale=0.3]{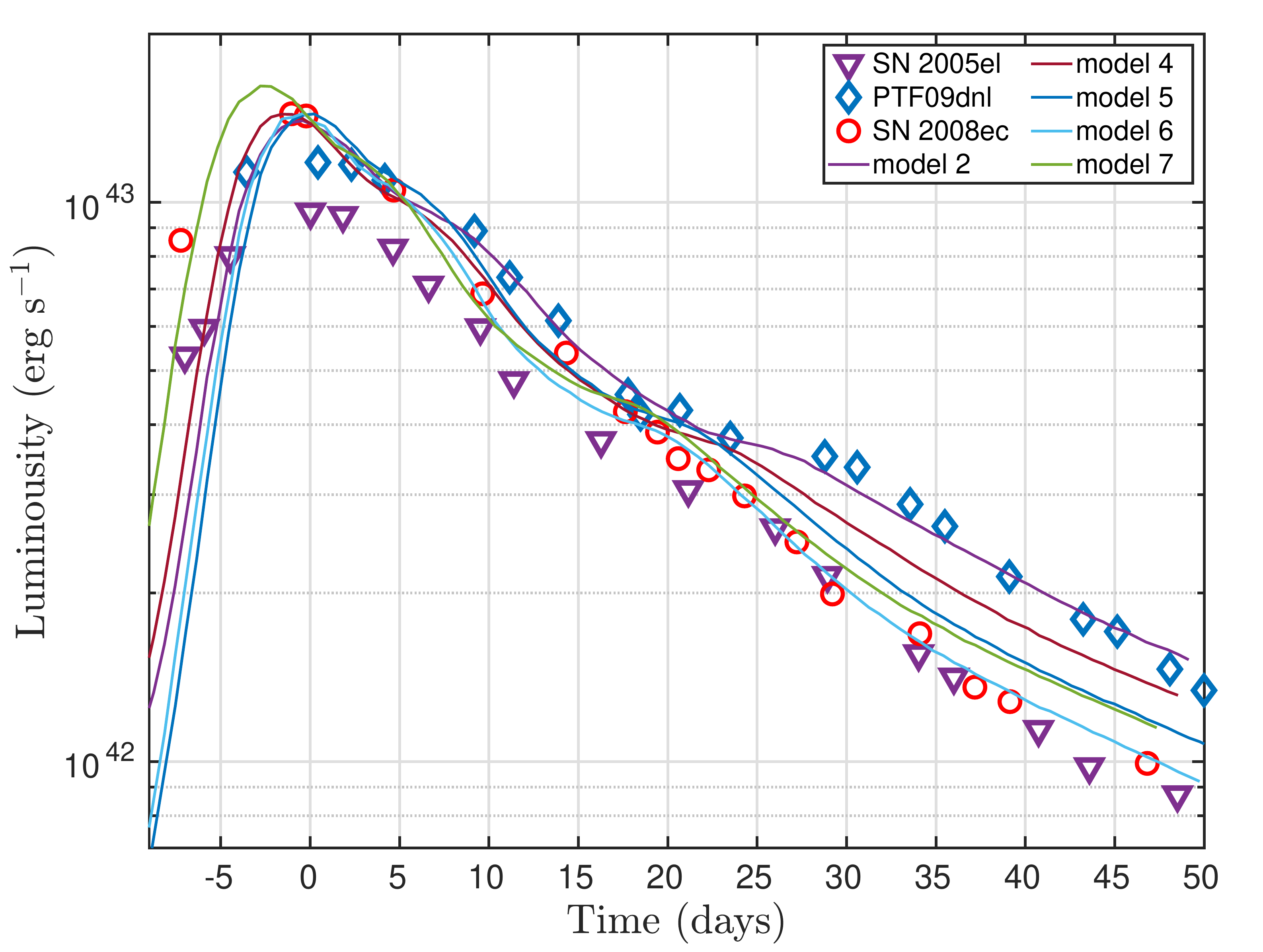}\includegraphics[scale=0.3]{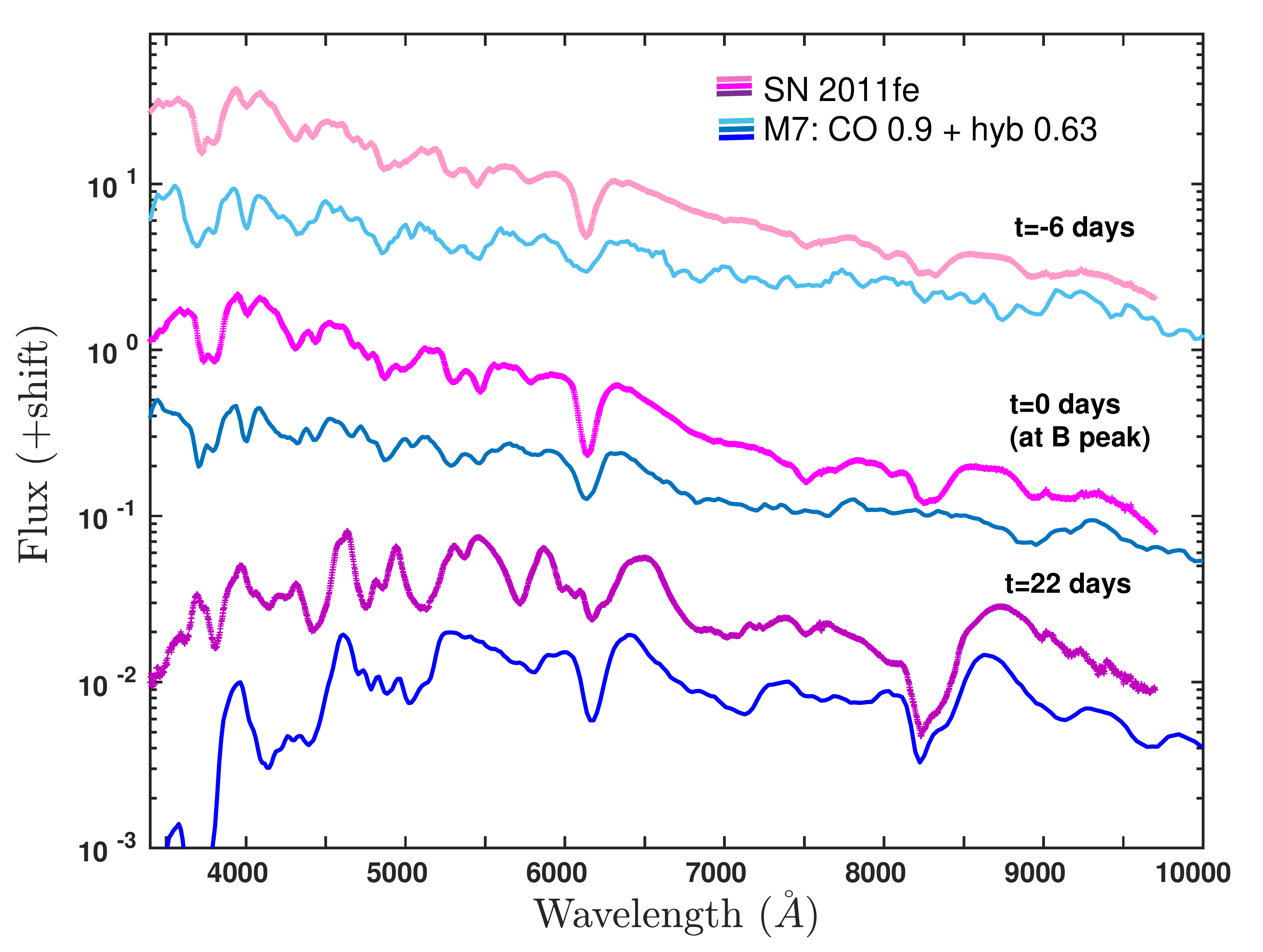}

\caption{\label{fig:lcs}{\footnotesize{}Comparison of the Bolometric and BVRI
light-curves and the spectra of the merger models and observations.
Left: Bolometric light curves for most of our models (the others show
qualitative similar evolution, and are not shown as to retain the
clarity of the figures) in comparison with inferred bolometric light
curves from \cite{Sca+14}. Note that the timing of the peak luminosity
postion of the theoretical light-curves were arbitrarily shifted as
to better show different models and their partial/ful fit to some
of the observed SNe. Top right: Comparison of SN 2008ec light curve
and our model 7 (CO 0.9 + hybrid 0.63). Bottom Right: Comparison of
the well observed spectra of SN 20011fe\cite{Per+13} with the spectra
from our model 7 (CO 0.9 + hybrid 0.63); the light curve evolution
of SN 2001fe somewhat differs that of sn 2008ec and we therefore compare
the spectra at days t=-5, t=0 and t=18 in the model with that at t=-6,
t=0 and t=22 for SN 20011fe observations. The model spectra are slightly
smoothed for clarity.}}
\end{figure}

Thousands of test-particles were inserted into each of our simulations
as to enable us to follow the detailed evolution of the material conditions
and apply a post-processing analysis of the results. We use the MESA-TORCH
module to post-process the nucleosynthetic burning of the material
throughout the explosion using a large nuclear network of 125 isotopes.
A summary of the resulting elemental products and in particular the
$^{56}{\rm Ni}$ production can be found in Table \ref{tab:initial-final}.
Interestingly, although our models produce a wide range of peak luminosities,
the different combinations of WD mergers give rise to a relatively
narrow range of $^{56}{\rm Ni}$ yields between 0.5-0.6 ${\rm M_{\odot}}$.
Using the resulting detailed compositional evolution output we then
follow the radiative-transfer evolution of the photons as to produce
detailed predictions for the light-curves and spectra expected from
each of the models. We make use of the publicly available SuperNu
code\cite{Wol+13}, using opacity mixing of 0.9\cite{Wol+13} (but
we checked the effect of using a lower, 0.5, value in some cases,
giving rise to up to $\sim0.15$ lower predicted peak B-magnitude)
and a group resolution of 625\cite{Wol+13} as to generate synthetic
light-curves and spectra for our models. In Fig. \ref{fig:lcs} we
show the bolometric light curves for our models in comparison to the
inferred bolometric light curves from {\small{}\cite{Sca+14}}. Our
models compare well with lower luminosity normal Ia SNe, but can not
reproduce brighter and slower evolving SNe. Our models give rise to
SNe with peak B (R)-luminosities in the range $-18.4\,-\,-19.2$ ($-18.5\,-\,-19.45$)
generally consistent with observed type Ia SNe (see Table \ref{tab:initial-final})

In Fig. we also prove a detailed example of the BVRI multi-band light-curves
from the result of our model 7 (0.9 ${\rm M_{\odot}}$ CO WD + 0.63
${\rm M_{\odot}}$ HeCO WD), compared with the relatively faint, but
normal type Ia SN 2008ec. In addition, we compare the spectra predicted
by model 7 with that of the best observed type Ia SN SN 2001fe. As
can be seen our predicted light-curves and spectra compare well with
the observed ones. Nevertheless, given the large phase-space of possible
explosions, the possible limitations of 2D models, and the many uncertainties
and assumptions in the construction of radiation-transfer models it
is not surprising that some differences can be identified, especially
in the B-band around 10-20 days post-peak, and in the differences
in the late secondary IR peak (the reproduction of which, however,
is notoriously difficult in radiative-transfer modeling, and likely
arises from the yet inaccurate assumptions and modeling made in the
codes\cite{Kas06}). Future improved models and even thicker sampling
of the phase-space of DWD-mergers might be able to further improve
the detailed quantitative reproduction of individual observed SNe. 

Note that our models can only reproduce the somewhat faster evolving
and somewhat fainter normal type Ia SNe, with all of them showing
${\rm M_{B}\gtrsim-19.2}$ and overall faster evolution (as expected
for fainter SNe), suggesting that slower and brighter type Ia SNe
arise from different progenitors. Most interestingly, mergers of massive
CO-CO WDs, not including hybrids, and with total combined masses larger
than$\sim1.9\,{\rm M_{\odot}}$ have been shown to produce slow evolving
relatively bright normal type Ia SNe, but could not produce the faster
and fainter ones (e.g. see refs. \cite{Kas+15,Pak17} for an overview;
note that \emph{peculiar,} SN 1991bg-like SNe, with $\Delta{\rm M_{B15}\gtrsim1.5}$
and ${\rm M_{B}\gtrsim-18}$, were suggested to be produced from violent
mergers of high mass, and high mass-ratio CO WDs\cite{pak+10}). In
other words mergers of massive CO WDs could provide a complementary
component for type Ia SNe progenitors, explaining the origin of slow
evolving type Ia SNe. As we discuss in the following, together He-CO
- CO mergers and massive-CO - massive-CO WD mergers can therefore
potentially explain the full range of type Ia SNe. We note that the
possibility of two type of progenitors might produce a systematic
non-continuous transition that could potentially be manifested observationally.
In that regard, we note in passing that the observed light-curve luminosity-width
relations suggest a possible bi-modal distribution of peak-luminosities
behavior (e.g. \cite{li+11}) close to the expected transition regions
of these parameters, possibly providing a clue for the suggested transition.
Moreover, recent studies suggested that some type Ia SNe show evidence
for a bi-modal velocity distribution of $^{56}$Ni, but only for the
fainter SNe\cite{Don+15}. In our models the two different detonations
lead to a bi-modal Ni production, and moreover, the disk material
and the material arising from the central CO WD have different velocities.
Together, these give rise to two velocity components with significantly
different amplitudes (thousands of km ${\rm s^{-1}}$ difference),
producing a bi-modal velocity distribution, while violent massive
CO-CO mergers should not. It is therefore possible that the observed
bi-modal velocity distribution and its brightness dependence is another
possible evidence for progenitors transition. 

\begin{figure}
\includegraphics[scale=0.3]{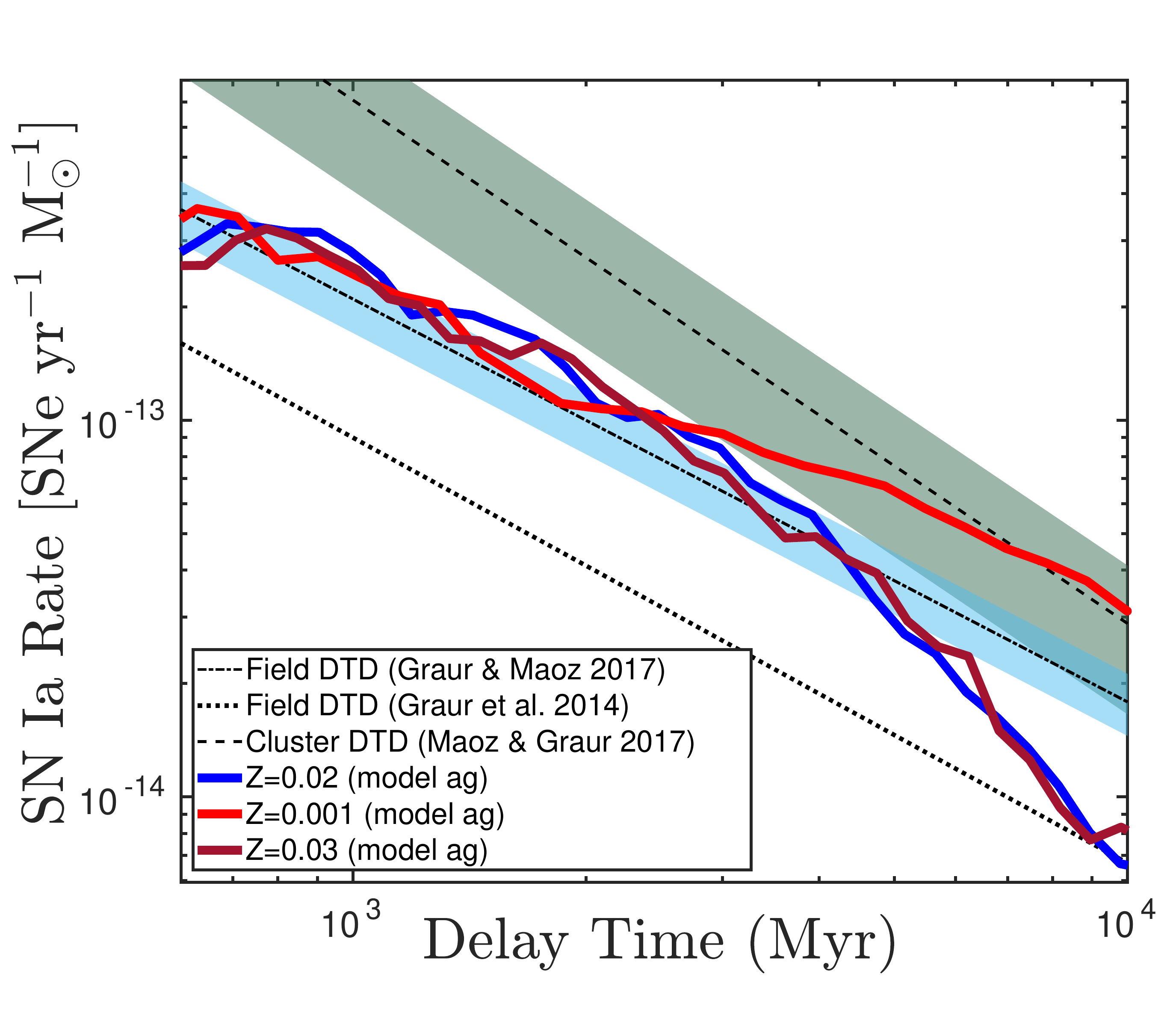}\includegraphics[scale=0.3]{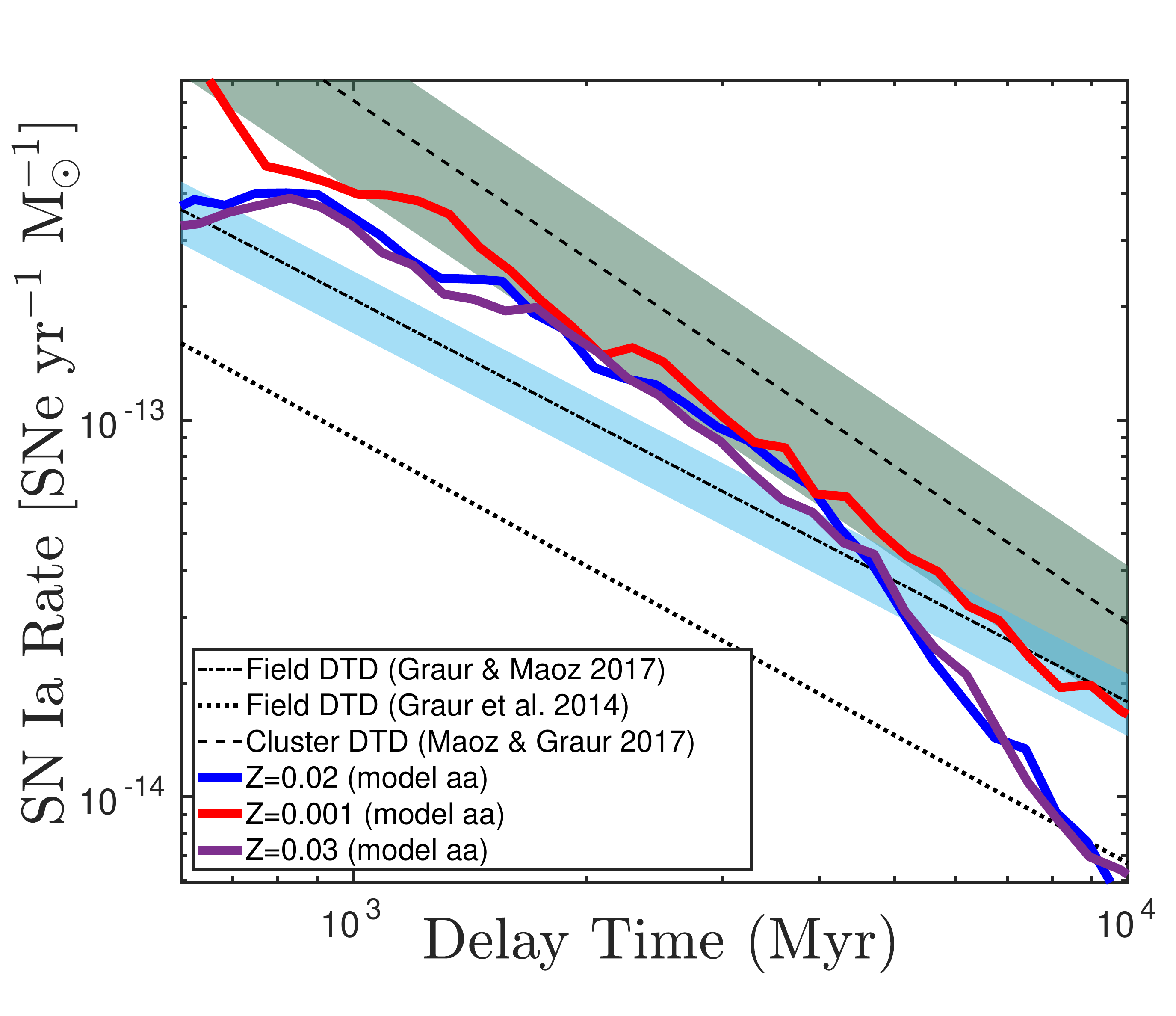}\caption{\label{fig:DTD} {\footnotesize{}Comparison of the modeled delay-time
distribution of SNe arising from our models with the DTDs inferred
from observations of field galaxies and cluster galaxies. Our models
consider all CO-primary + hybrid-secondary WD mergers with a total
combined mass $>1.15$ ${\rm M_{\odot}}$ as well as all CO+CO WD
mergers with a combined total mass larger than $1.9$ M$_{\odot}$.
Models for several different metalicities are shown (${\rm Z=0.02,\,0.001}$and
$0.03$). The left plot corresponds to $\alpha\gamma$ models and
the right plot shows the results from the $\alpha\alpha$ binary population
synthesis models. See main text for further details. }}
\end{figure}

Given that our models can generally reproduce the detailed light-curves
and spectra of normal but fainter (${\rm M_{B}}\sim-18.4-19.2$; ${\rm M_{R}\sim-18.5-19.4}$)
type Ia SNe, and their diversity, we now consider their expected rates
and delay-time distribution. In order to study these issues we apply
binary population synthesis models using the SeBa code \cite{Por+96,Too+12}.
Given our successful models, we identify standard type Ia progenitors
a pairs of WD merging over a Hubble time with a primary CO WD disrupting
a hybrid HeCO WD and a total combined mass $>1.15$ ${\rm M_{\odot}}$
(simulations of mergers with total masses lower than this value produced
faint SNe, not resembling normal type Ia SNe, which are explored elsewhere).
Since previous models have shown that mergers of two CO WDs can reproduce
slow evolving, bright type Ia SNe when the combined mass was $\gtrsim1.9$
${\rm M_{\odot}}$ we also identify such cases with type Ia SNe (but
we do not consider models of dynamical detonation such as \cite{She+18}).
We consider two typical models, $\alpha\alpha$ and $\alpha\gamma$,
which differ in their assumptions regarding common envelope evolution
in binaries. Our detailed initial conditions and assumptions regarding
binary stellar evolution are described in detail in the Methods section,
and the composition of our hybrid HeCO WDs is determined following
our detailed binary evolution models in \cite{Zen+18a}. The main
results are shown in Fig. \ref{fig:DTD}. As can be seen our models
not only reproduce the overall inferred rates of type Ia SNe, but
are also consistent with the inferred\cite{2014ARA&A..52..107M,Gra+14,Mao+17}
DTD of these SNe. While HeCO-CO WD mergers contribute the majority
of SNe ($60-70\%;$ depending on the model), and produce fainter,
faster evolving SNe, mergers of massive double CO-WDs (typically both
above $0.9$ ${\rm M_{\odot}})$ also provide a critical contribution
($\sim30-40\%$) and complement the models explored here, explaining
brighter, slower SNe (and possibly the 91bg-like ultra faint and fast
evolving SNe from violent mergers of comparable mass CO WDs\cite{pakmoretal10}).

type Ia SNe play a key-role in the production of the elements in the
universe, serve as standard-candles for cosmological distance-measurements
and affect the evolution of galaxies and star-formation through feedback.
Providing a coherent model for the origins of such SNe as suggested
here is therefore paramount for our understanding of the chemical
and dynamical evolution of galaxies. Moreover, such a model can shed
light on the systematics involved in the properties of SNe and their
dependence on the properties of the progenitors and the cosmic evolution
(e.g. through the dependence on metallicity; see the difference in
DTD between high and low metallicity environments shown in Fig. \ref{fig:DTD}),
and can therefore potentially inform our understanding of the systematics
involved in the calibration of type Ia SNe as standard candles, and
their implications for the measurements of the fundamental parameters
of the universe and its constituents. 

\section{Methods}

\subsection{Hydrodynamical simulations, post-processing and initial configurations}

\label{subsec:FLASH+supernu}

All of our merger models were simulated using the publicly available
FLASH v4.5 code \cite{2000ApJS..131..273F}, and following the same
methods employed by us in \cite{Zen+18b,Zen+19}, which we briefly
review here again. The simulations were done using the unsplit ${\rm PPM}$
solver of FLASH in ${\rm 2D}$ axisymmetric cylindrical coordinates
on a grid of size ${\rm 1\times1\left[10^{10}cm\right]}$ using adaptive
mesh refinement. We follow similar approaches as described in other
works on thermonuclear SNe (e.g. \cite{Mea+09}). Detonations are
handled by the reactive hydrodynamics solver in FLASH without the
need for a front tracker, which is possible since unresolved Chapman--Jouguet
(CJ) detonations retain the correct jump conditions and propagation
speeds. Numerical stability is maintained by preventing nuclear burning
within the shock. This is necessary because shocks are artificially
spread out over a few zones by the ${\rm PPM}$ hydrodynamics solver,
which can lead to nonphysical burning within shocks that can destabilize
the burning front \cite{1989BAAS...21.1209F}.

The properties of the WDs considered in our models are obtained through
detailed stellar evolution models using the MESA code \cite{2011ApJS..192....3P,2015ApJS..220...15P}.
In all cases we considered only Solar metallicity stellar progenitors.
The primary CO WDs are produced from the regular evolution of single
stars, that eventually produce WDs composed of $\sim$50$\%$ carbon
and $\sim$50$\%$ oxygen. The hybrid WDs, containing both CO and
He are derived from detailed \emph{binary} evolution in MESA, as described
in \cite{Zen+18a}. Before the beginning of the simulation the central
WD is relaxed following the procedure in \cite{deV+14}, by letting
the WD evolve in isolation and repeatedly damping any residual velocities
during the few dynamical timescales evolution; finally afterwards
we also make use of the damp\_method in module FLASH. 

Following the results of SPH simulations of DWD-mergers we assume
the disk composition is fully mixed; the specific fractional composition
follows the composition of hybrid WDs as obtained from out full stellar
evolution models (using the MESA code; see \cite{Zen+18a}). The disk
structure follows the same methods as used by us earlier \cite{Zen+18b},
where the appropriate velocities in the disk are accounted for, given
the central potential of the WD. Note that we account for the self-gravity
of the disk and self-consistently derived the structure and appropriate
equation of state throughout the disk, through an iterative convergence
method (see \cite{Zen+18b} for details), which corrects the appropriate
equation of state (EOS) used in the different parts of the disk, and
the velocities of the material, before initializing the full FLASH
simulation. 

The nuclear network used is the FLASH $\alpha-$chain network of 19
isotopes \cite{1989BAAS...21.1209F}. This network can adequately
capture the energy generated during the nuclear burning \cite{2000ApJS..126..501T}.
In order to follow the post-process analysis of the detailed nucleosynthetic
processes and yields we made use of $4000-20000$ tracer particles
that track the radius, velocity, density, and temperature and are
evenly spaced every ${\rm 2\times10^{7}cm}$ throughout the WD-debris
disk. 

We made multiple simulations with increased resolution until convergence
was reached in the nuclear burning. We found a resolution of $1-10$
km to be sufficient for convergence of up to 10$\%$ in nuclear energy
produced. Gravity was included as a multipole expansion of up to multipole
$l=12-40$ using the new FLASH multipole solver, to which we added
a point-mass gravitational potential to account for gravity of the
NS. We simulated the viscous term by using the viscosity unit in Flash,
employing a \cite{1973A&A....24..337S} parameterization ${\rm \nu_{\alpha}=\alpha C_{s}^{2}/\Omega_{Kepler}}$,
where ${\rm \Omega_{Kepler}}$ is the Keplerian frequency and ${\rm C_{s}}$
is the sound speed. The contributions of both nuclear reaction and
neutrino cooling \cite{1989ApJ...346..847C,1991ApJ...376..234H} are
included in the internal energy calculations, and the Navier-Stokes
equations are solved with source terms due to gravity, shear viscosity
and the nuclear reactions. 

The EOS used in our simulations is the detailed Helmoholz EOS employed
in FLASH (\cite{2000ApJS..126..501T}). This EOS includes contributions
from partially degenerate electrons and positrons, radiation, and
non-degenerate ions. It uses a look-up table scheme for high performance.
The most important aspect of the Helmholtz EOS is its ability to handle
thermodynamic states where radiation dominates, and under conditions
of very high pressure.

All our simulations were run down to a resolution of 8 km. For five
cases we made further higher resolution runs down to 5 km, and verified
that the identified detonations occur at the same positions (the same
corresponding simulation cells) and at the same times (up to a difference
of a few milliseconds). We also verified that the overall evolution
as well as the energetics, angular momenta and elemental yields were
conserved when comparing the lower and higher resolution runs. We
also verified that the snapshots just before the detonations present
increased burning, and increased densities and velocities leading
the detonation. 

In order to prevent the production of artificial unrealistic early
detonation that may arise from insufficient numerical resolution,
we applied a limiter approach following \cite{2013ApJ...778L..37K}.
This burning limiter suppresses artificial ignitions in low resolutions,
but does not affect the process of ignition when it is resolved \cite{2013ApJ...778L..37K};
our resolution is also comparable with that suggested by ref. \cite{2013ApJ...778L..37K}
to be sufficient for resolving the CO ignition region. We note that
Dan et al.\cite{Dan+15}, modeled a merger of CO WDs with Helium-rich
CO-WDs, but only one of their models had a realistic hybrid-WD composition.
Their runs employed a 13 elements network, and did not use the limiter
approach. Their model resulted in a much less energetic explosion
producing only $\sim0.3\,M_{\odot}$ of $^{56}{\rm Ni}.$ For comparison
we rerun a similar model, where we did not employ the burning limiter
(though still with 19 elements network). In this case we obtained
only $\sim0.4$ ${\rm M_{\odot}}$ of ${\rm ^{56}{\rm Ni}}$, more
similar to the results of \cite{Dan+15}, suggesting the non trivial
role of the burn-limiter criterion, and the reason that model do not
produce a typical SN Ia-like explosion. 

We find that our He detonations occur at external, less resolved regions
as discussed in the main text. Nevertheless, even in these regions
our resolution is below $\sim80$ km, comparable or even better than
the $\sim100$ km size typically found necessary for resolving succesful
He detonations under comparable conditions\cite{2013ApJ...774..137M,Hol+13}.

Following the FLASH runs we make use of the detailed histories of
the tracer particles density and temperature to be post-processed
with MESA (version 8118) one zone burner \cite{2015ApJS..220...15P}.
We employed a 125-isotope network that includes neutrons, and composite
reactions from JINA\textquoteright s REACLIB \cite{2010ApJS..189..240C}.
Overall we find that the results from the larger network employed
in the post-process analysis show comparable, but somewhat less efficient
nuclear burning.

The tracer particles and their compositional data from the post-processing
step is provided as input for our radiative-transfer modeling using
the openly available SuperNu code. SuperNu is a multi-dimensional
radiative-transfer code, which can provide detailed light-curves and
spectra as observed from different directions. SuperNu simulates time-dependent
radiation transport in local thermodynamic equilibrium with matter.
It makes use of Implicit Monte Carlo and Discrete Diffusion Monte
Carlo methods for static or homologously expanding spatial grids.
It is used for post-processing analysis, and the radiation field affects
material temperature, but does not affect the motion of the fluid
which is only modeled in the previous hydrodynamics stage using FLASH.
Detailed description and comparison of the code with other codes can
be found in Refs. \cite{Wol+13,wollaegervanrossum14}. Although radiative
transfer codes for SN modeling have been significantly developed over
the past decade, they still encounter many challenges and difficulties,
and although different codes do give rise to qualitatively similar
light-curve and spectra, there are still non-negligible quantitative
differences arising from the different implementations and underlying
assumptions; and therefore the level of accuracy of the resulting
light-curve and spectra should be considered accordingly. In particular
the structure of the secondary peak in the NIR bands are notoriously
sensitive to the atomic lines used and the level of mixing assumed\cite{Kas06}. 

\subsection{Merger models}

Table \ref{tab:initial-final} provides the details for all of the
models explored in this study, and their general physical and observational
outcomes. We generally explored a wide range of mass combinations
of Co and HeCO WDs. The computational expense limits the number of
the models runs; a more detailed grid would be explored in future
studies. 

\begin{table*}[t]
\begin{tabular}{|c|c|c|c|c|c|c|c|c|}
\hline 
{\small{}\# } & {\small{}${\rm M^{Hyb}}{\rm (M_{He4}[M_{\odot}]})$ } & {\small{}${\rm M_{CO}[M_{\odot}]}$ } & {\small{}${\rm M_{tot}[M_{\odot}]}$ } & {\small{}${\rm R_{d}}$(${\rm R_{t}})$} & {\small{}${\rm M_{Ni56}[M_{\odot}]}$ } & {\small{}${\rm E_{K}/10^{51}[erg]}$} & {\small{}${\rm B}_{{\rm peak}}$} & {\small{}${\rm R}_{{\rm peak}}$}\tabularnewline
\hline 
\hline 
{\small{}1} & {\small{}0.53 (0.074)} & {\small{}$0.7$ } & {\small{}$1.23$ } & {\small{}$1$} & {\small{}$0.512$ } & {\small{}$0.484$ } & {\small{}$-18.40$} & {\small{}$-18.73$}\tabularnewline
\hline 
{\small{}2} & {\small{}0.53 (0.074)} & {\small{}$0.75$ } & {\small{}$1.28$ } & {\small{}$1$} & {\small{}$0.533$ } & {\small{}$0.412$ } & {\small{}$-18.49$} & {\small{}$-18.52$}\tabularnewline
\hline 
{\small{}3} & {\small{}0.53 (0.074)} & {\small{}$0.8$ } & {\small{}$1.33$ } & {\small{}$1$} & {\small{}$0.530$ } & {\small{}$0.511$ } & {\small{}$-18.69$} & {\small{}$-18.73$}\tabularnewline
\hline 
{\small{}4} & {\small{}0.53 (0.074)} & {\small{}$0.9$ } & {\small{}$1.43$ } & {\small{}$1$} & {\small{}$0.549$ } & {\small{}$0.527$ } & {\small{}$-18.67$} & {\small{}$-18.72$}\tabularnewline
\hline 
{\small{}5} & {\small{}0.63 (0.03)} & {\small{}$0.75$ } & {\small{}$1.38$ } & {\small{}$1$} & {\small{}$0.538$ } & {\small{}$0.501$ } & {\small{}$-18.57$} & {\small{}$-18.52$}\tabularnewline
\hline 
{\small{}6} & {\small{}0.63 (0.03)} & {\small{}$0.8$ } & {\small{}$1.43$ } & {\small{}$1$} & {\small{}$0.556$ } & {\small{}$0.541$ } & {\small{}$-18.48$} & {\small{}$-19.32$}\tabularnewline
\hline 
{\small{}7} & {\small{}0.63 (0.03)} & {\small{}$0.9$ } & {\small{}$1.53$ } & {\small{}$1$} & {\small{}$0.562$ } & {\small{}$0.594$ } & {\small{}$-18.52$} & {\small{}$-18.54$}\tabularnewline
\hline 
{\small{}8} & {\small{}0.63 (0.03)} & {\small{}$1.0$ } & {\small{}$1.63$ } & {\small{}$0.8$} & {\small{}$0.564$ } & {\small{}$0.608$ } & {\small{}$-18.63$} & {\small{}$-19.07$}\tabularnewline
\hline 
{\small{}9} & {\small{}0.68 (0.015)} & {\small{}$1$ } & {\small{}$1.68$ } & {\small{}$1$} & {\small{}$0.569$ } & {\small{}$0.615$ } & {\small{}$-18.71$} & {\small{}$-19.08$}\tabularnewline
\hline 
{\small{}10} & {\small{}0.68 (0.015)} & {\small{}$1$ } & {\small{}$1.68$ } & {\small{}$0.8$} & {\small{}$0.574$ } & {\small{}$0.621$ } & {\small{}$-18.80$} & {\small{}$-19.26$}\tabularnewline
\hline 
{\small{}11} & {\small{}0.74 (0.01)} & {\small{}$0.9$ } & {\small{}$1.64$ } & {\small{}$0.8$} & {\small{}$0.589$ } & {\small{}$0.635$ } & {\small{}$-19.05$} & {\small{}$-19.34$}\tabularnewline
\hline 
{\small{}12} & {\small{}0.74 (0.01)} & {\small{}$1.0$ } & {\small{}$1.74$ } & {\small{}$1$} & {\small{}$0.592$ } & {\small{}$0.645$ } & {\small{}$-19.16$} & {\small{}$-19.44$}\tabularnewline
\hline 
\end{tabular}

\caption{\label{tab:initial-final} The merger models explored in this study,
covering a wide range of CO and He-CO WD combinations, providing the
most detailed sampling of WD merger models to-date with synthetic
light-curve and spectra. The columns correspond to the model number
(1), The mass (and Helium mass) in the hybrid WDs (2), The CO WD mass
(3); ${\rm R_{d}}$ in units of the tidal radius (${\rm R_{t}})$
(4); the total mass of the merging WDs (5); the amount of $^{56}{\rm Ni}$
produced (6); the total kinetic energy (7); the B-peak luminousity
(8); the R-peak luminousity (9).}
\end{table*}

\subsection{Binary population synthesis}

\label{subsec:bps}

The formation and evolution of interacting binaries producing WD-WD
mergers is modeled with the binary population synthesis (BPS) code
\texttt{SeBa} \cite{Por+96,Too+12}, following the same procedures
and general assumptions used by us in previous studies\cite{Zen+18a,Ton+18}.
Here we briefly review them again.

\texttt{SeBa} is a code for fast-modeling of binary evolution based
on parameterized stellar evolution, including processes such as mass
transfer episodes, common-envelope evolution and stellar winds. Using
\texttt{SeBa} we generate a large population of binaries on the zero-age
MS, model their subsequent evolution, and extract those that produce
mergers of two WDs. We identify hybrid WDs in the BPS models using
the results of the detailed MESA stellar evolution of hybrid HeCO
WDs which we previously explored\cite{Zen+18a}. 

\cite{Too+14} have already shown that the main sources of differences
between different BPS codes is due to the choice of input physics
and initial conditions. Here we focus only on a specific set of choices.
We construct two models that differ with respect to the assumptions
regarding the common-envelope phase. The common-envelope phase can
occur during a short epoch in the evolution of a binary system when
both stars share a common-envelope. Despite its strong effect on the
binary orbit, common-envelope evolution is poorly understood \cite[for a review]{Iva13}.
We replicate model $\alpha\alpha$ and model $\alpha\gamma$ from
\cite{Too17}, where the prior is based on the classical energy balance
during the common-envelope phase, whereas the latter is based on a
balance of angular momentum. Note that the $\alpha\alpha$ model is
the typical approach to model common-envelope evolution in BPS, while
the $\alpha\gamma$ model was constructed to better fit the mass ratios
of observed DWDs\cite{2001A&A...365..491N}.

We consider a classical set-up for BPS calculations, initial conditions
and assumptions: (1) The primary masses are drawn from a Kroupa IMF
\cite{Kro93} with masses in the range between $0.1-100$ ${\rm M_{\odot}}$.
(2) The secondary masses are drawn from a uniform mass ratio distribution
with $0<q\equiv M_{2}/M_{1}<1$ \cite{Rag+10,Duc+13,DeR14}. (3) The
orbital separations $a$ follow a uniform distribution in $log(a)$
\cite{Abt83} (4) The initial eccentricities $e$ follow a thermal
distribution \cite{Heg75}. (5) A binary fraction $\mathcal{B}$ of
75$\%$ which is appropriate for A/B-type primaries \cite{Rag+10,Duc+13}. 

\subsection{Shock compression }

\label{subsec:shock}

The production of $^{56}{\rm Ni}$ typically require the nuclear burning
of high density material (typically $>10^{7}$g cm$^{-3}$). However,
such high densities exist only in massive WDs with typical masses
close or above $0.9$ ${\rm M_{\odot}}$. Nevertheless, low-mass WDs
could also give rise to such high densities through shock compression.
For example, even low mass $\sim0.64$ ${\rm M_{\odot}}$ gave rise
to high $^{56}{\rm Ni}$ yields in simulations of direct physical
collisions between WDs\cite{2013ApJ...778L..37K,Pap+16}. In that
case the shock from the collision itself compressed the WD material,
leading to its detonation and the abundant production of $^{56}{\rm Ni}$.
The shock energy in the collision case originated from the gravitational
binding energy of the WD pair, translated into kinetic energy and
giving rise to high supersonic collision velocity and the compressive
shock. 

\begin{figure}
\includegraphics[scale=0.3]{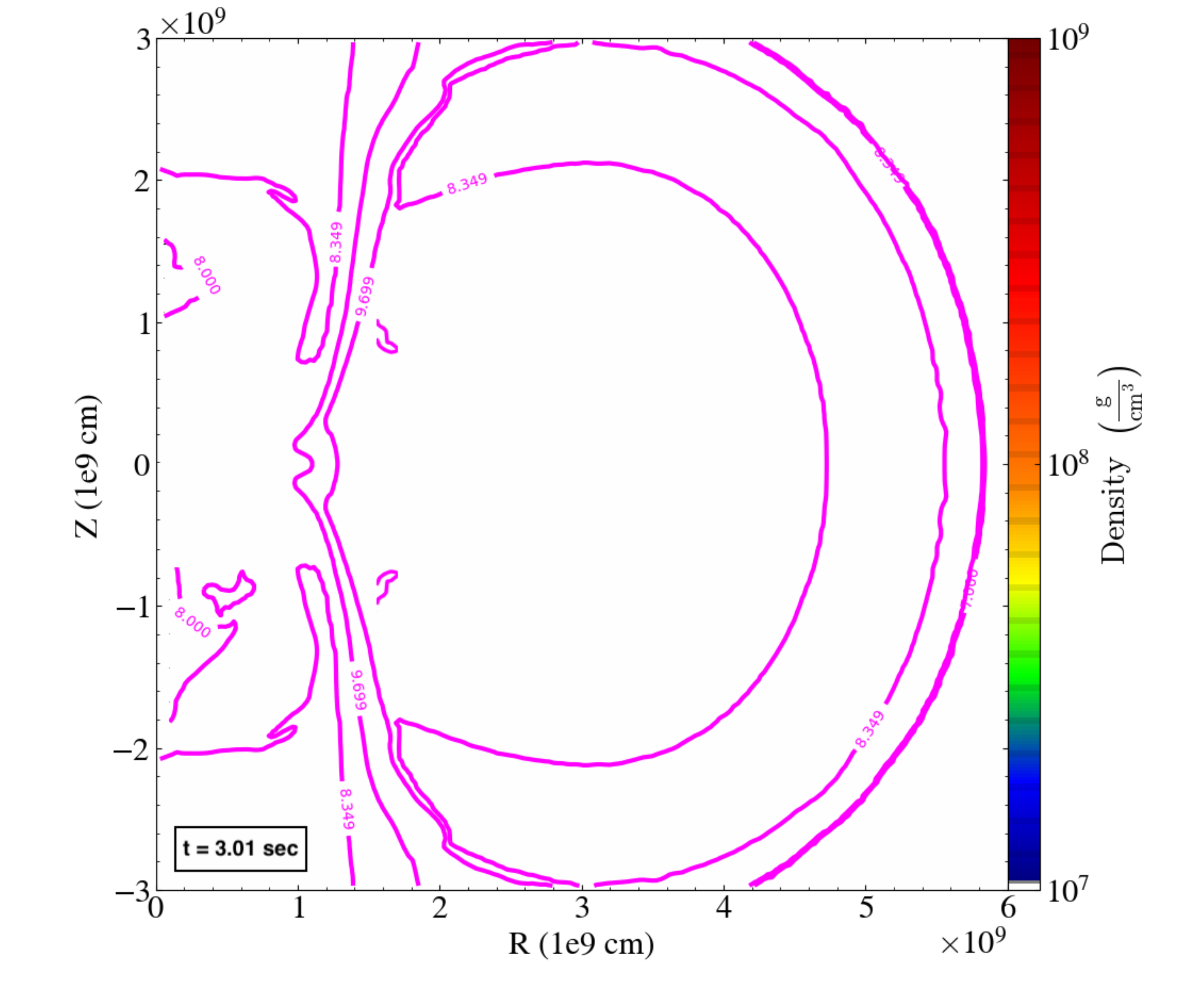}\includegraphics[scale=0.3]{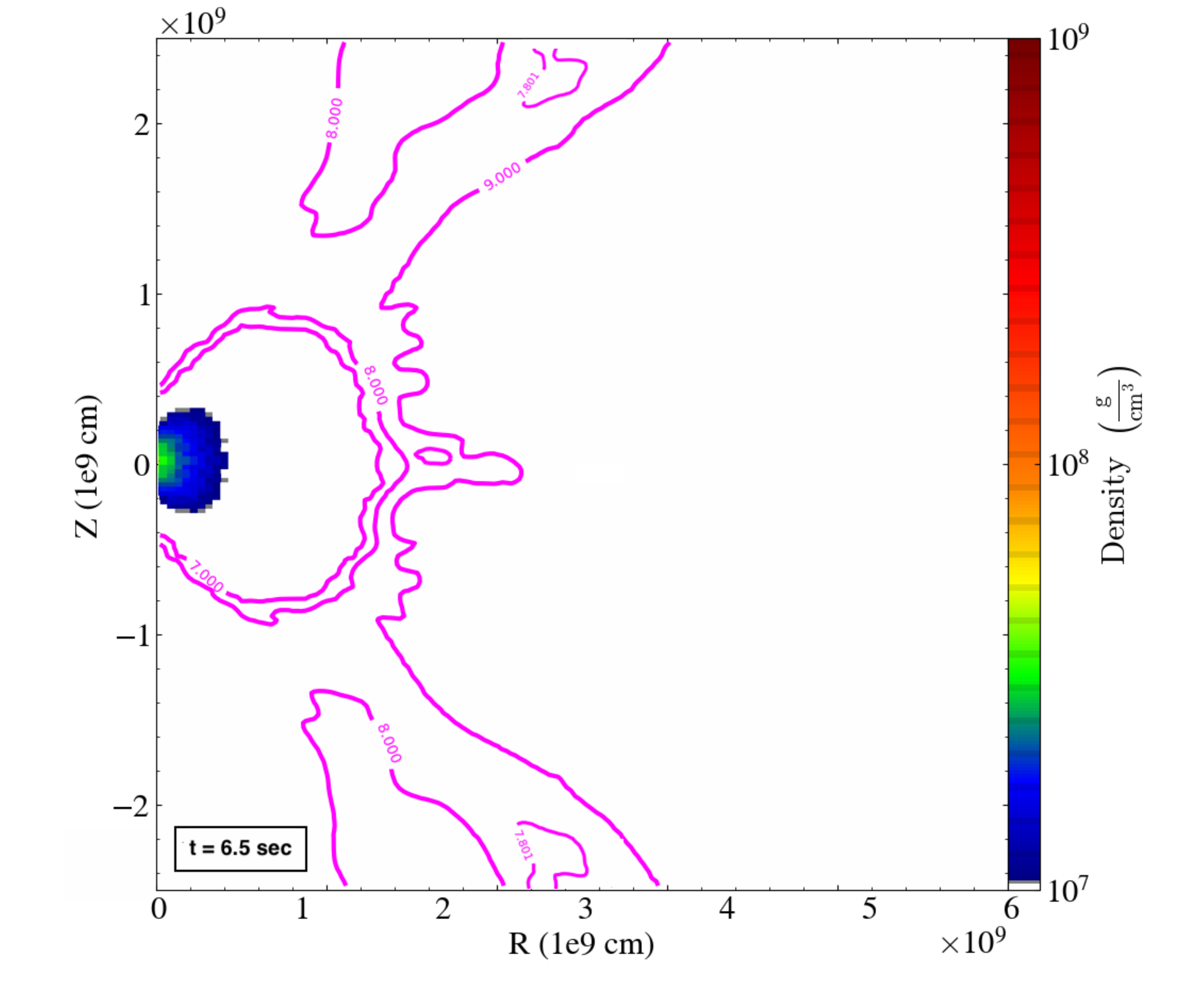}\caption{\label{fig:high-density}{\footnotesize{}The distribution of high
density material (only material with $>10^{7}$ cm$^{-3}$ is shown)
before the first and second detonations. Just before the first, He-rich
detonation (left panel) none of the material of the 0.8 ${\rm M_{\odot}}$CO
WD attained high densities. However, following the shock compression
due to the He-detonaion, and just before the second CO detonation
the central parts of the CO WD were compressed to very high densities
(up to $6-7\times10^{7}$ g cm$^{-3}$), allowing for the CO detonation
and the efficient nuclear burning of most of the WD up to $^{56}{\rm Ni}$.
Ths pink lines show the temperature contours.}}
\end{figure}

In the case of the CO-HeCO merger we identify an initial detonation
in the He-rich material accreting on the CO WD, producing a few $0.01$${\rm M_{\odot}}$of
$^{56}{\rm Ni}$, and thereby releasing nuclear energy which is comparable
or even higher than the energy deposited during a physical collision
between two WDs. The He-rich detonation can therefore serve as to
produce a compressive shock that compresses the CO WD. In particular
we find that even $0.7$${\rm M_{\odot}}$ WDs can be sufficiently
compressed as to compress a large fraction of the WD material up to
the critical density ($>10^{7}$ g cm$^{-3}$) and allowing for for
abundant $^{56}{\rm Ni}$ production following the second detonation
of the core of the CO WD. As can be seen in Fig. \ref{fig:high-density}
this is indeed the case in our simulations. Shown is the density distribution
of material above $10^{7}$ g cm$^{-2}$ for a $0.8$ M$_{\odot}$
CO WD, just before the first He-rich detonation and then just before
the second CO detonation. As can be seen effectively not even the
central core of the WD had such high densities before the He-detonation,
but following the He-detonation the CO WD was shock compressed such
that most of its material attained high densities, even as high as
a few times $10^{7}$ g cm$^{-3}$. 

Indeed, the central densities of $0.7$ CO WDs are only a factor of
$\sim2-6$ (from the inner regions up to 70$\%$ of the total mass)
lower than the critical density. The specific volume ratio between
the pre-shocked material and the shocked material is a function of
pressure and ${\rm \gamma}$ (the adiabatic exponent in the equation
of state), and the limiting density across the shock wave is then
${\rm \rho_{a}/\rho_{0}=\gamma+1/\gamma-1}$. For degenerate material
with ${\rm \gamma=4/3}$ we therefore expect the limiting density
to be $7$, down to 4 for $\gamma=5/3$. In our case the CO core is
highly degenerate, and it is therefore possible for the He-detonation
shock to provide the necessary compression. This also suggest that
$0.7$ ${\rm M_{\odot}}$ CO WDs would give rise to a lower limit
for the expected WDs that may produce a $^{56}{\rm Ni}$ rich explosion,
consistent with our results showing no normal type Ia SNe (and $^{56}{\rm Ni}$
rich $>0.45\text{{\rm M\ensuremath{_{\odot}})}}$ from progenitors
below 0.7 ${\rm M_{\odot}}$. Do note that our 2D models are axially
symmetric, by construction, and that they may therefore potentially
introduce artificially stronger compression than might be produced
by an initially localized detonation. It is therefore possible that
the actual lower limit for a $^{56}{\rm Ni}$ rich explosions would
be at a somewhat more massive WD regime than the 0.7 ${\rm M_{\odot}}$
WDs limit found in our 2D models. 



\bibliographystyle{naturemag}

\end{document}